\documentclass[12pt,a4paper]{article}

\usepackage{amsmath}
\usepackage{amssymb}
\usepackage{epsfig}
\usepackage{graphicx}
\usepackage{cite}

\newcommand{\capdef}{}
\newcommand{\mycaption}[2][\capdef]{\renewcommand{\capdef}{#2}%
       \caption[#1]{{\footnotesize #2}}}
\makeatletter
\renewcommand{\fnum@table}{\textbf{\tablename~\thetable}}
\renewcommand{\fnum@figure}{\textbf{\figurename~\thefigure}}
\makeatother

\newcounter{myenumi}

\renewcommand{\themyenumi}{\roman{myenumi}}
{\end{list}}

\newlength{\myem}
\newcounter{mysubequation}[equation]

\makeatletter
\renewcommand{\section}{\@startsection{section}{1}{0em}{-\baselineskip}%
{\baselineskip}{\normalfont\large\bfseries}}
\renewcommand{\subsection}%
{\@startsection{subsection}{2}{0em}{-0.7\baselineskip}%
{0.7\baselineskip}{\normalfont\bfseries}}
\makeatother

\newcommand{\bea}{\begin{eqnarray*}}
\newcommand{\eea}{\end{eqnarray*}}

\newcommand{\deltacp}{\delta_\mathrm{CP}}

\newcommand{\dm}[1]{{\Delta m^2_{#1}}}

\newcommand{\ie}{{\it i.e.}}

\newcommand{\eg}{{\it e.g.}}

\newcommand{\cf}{{\it cf.}}

\newcommand{\eq}{Eq.}
\newcommand{\eqs}{Eqs.}

\newcommand{\fig}{Figure}
\newcommand{\Fig}{Figure}
\newcommand{\figs}{Figures}
\newcommand{\Figs}{Figures}
\newcommand{\Ref}{Ref.}
\newcommand{\Refs}{Refs.}
\newcommand{\Sec}{Section}
\newcommand{\Secs}{Sections}
\newcommand{\App}{Appendix}

\newcommand{\Tab}{Table}

\newtheorem{definition}{Definition}

\newcommand{\JHFSK}{\mbox{\sf JPARC-SK}}
\newcommand{\NuMI}{\mbox{\sf NuMI}}
\newcommand{\minos}{\mbox{\sf MINOS}}
\newcommand{\icarus}{\mbox{\sf ICARUS}}
\newcommand{\opera}{\mbox{\sf OPERA}}
\newcommand{\CHOOZII}{\mbox{\sf D-Chooz}}
\newcommand{\DChooz}{\mbox{\sf Double-Chooz}}
\newcommand{\ReactorII}{\mbox{\sf Reactor-II}}

\newcommand{\stheta}{\sin^22\theta_{13}}
\newcommand{\ldm}{\Delta m_{31}^2}
\newcommand{\sdm}{\Delta m_{21}^2}
\newcommand{\equ}[1]{\eq~(\ref{equ:#1})}
\newcommand{\figu}[1]{\fig~\ref{fig:#1}}
\newcommand{\bi}{\begin{itemize}}
\newcommand{\ei}{\end{itemize}}

\begin{document}
%%%%%%%%%%%%%%%%%%%%%%%%%%%%%%%%%%%%%%%%%%%%%%%%%%%%%%%%%%%%%%%%%%%%%
%%%%                     Title-page                              %%%%
%%%%%%%%%%%%%%%%%%%%%%%%%%%%%%%%%%%%%%%%%%%%%%%%%%%%%%%%%%%%%%%%%%%%%

\begin{titlepage}

% the footnote symbols are only redefined for the title page !
\renewcommand{\thefootnote}{\alph{footnote}}

\vspace*{-3.cm}
\begin{flushright}
TUM-HEP-545/04\\
MPP-2004-28\\
%hep-ph/
\end{flushright}

\vspace*{0.5cm}

\renewcommand{\thefootnote}{\fnsymbol{footnote}}
\setcounter{footnote}{-1}

{\begin{center}
{\Large\bf Prospects of accelerator and reactor neutrino\\[2mm] 
oscillation experiments for the coming ten years}
\end{center}}
\renewcommand{\thefootnote}{\alph{footnote}}

\vspace*{.8cm}
%\vspace*{.3cm}
{\begin{center} {\large{\sc
                P.~Huber\footnote[1]{\makebox[1.cm]{Email:}
                phuber@ph.tum.de},~
                M.~Lindner\footnote[2]{\makebox[1.cm]{Email:}
                lindner@ph.tum.de},~
                M.~Rolinec\footnote[3]{\makebox[1.cm]{Email:}
                rolinec@ph.tum.de},\\
                T.~Schwetz\footnote[4]{\makebox[1.cm]{Email:}
                schwetz@ph.tum.de},~and~
                W.~Winter\footnote[5]{\makebox[1.cm]{Email:}
                wwinter@ph.tum.de}
                }}
\end{center}}
\vspace*{0cm}
{\it
\begin{center}

\footnotemark[1]${}^,$\footnotemark[2]${}^,$\footnotemark[3]${}^,$\footnotemark[4]${}^,$\footnotemark[5]%
       Physik--Department, Technische Universit\"at M\"unchen,\\
       James--Franck--Strasse, D--85748 Garching, Germany

\footnotemark[1]%
       Max-Planck-Institut f\"ur Physik, Postfach 401212,
       D--80805 M\"unchen, Germany

\end{center}}

\vspace*{1cm}

\begin{abstract}
   We analyze the physics potential of long baseline neutrino
   oscillation experiments planned for the coming ten years, where the
   main focus is the sensitivity limit to the small mixing angle
   $\theta_{13}$. The discussed experiments include the conventional
   beam experiments MINOS, ICARUS, and OPERA, which are under
   construction, the planned superbeam experiments J-PARC to
   Super-Kamiokande and NuMI off-axis, as well as new reactor
   experiments with near and far detectors, represented by the
   Double-Chooz project. We perform a complete numerical simulation
   including systematics, correlations, and degeneracies on an equal
   footing for all experiments using the GLoBES software. After
   discussing the improvement of our knowledge on the atmospheric
   parameters $\theta_{23}$ and $\Delta m^2_{31}$ by these experiments,
   we investigate the potential to determine $\theta_{13}$ within the 
   next ten years in detail. Furthermore, we show that under optimistic
   assumptions and for $\theta_{13}$ close to the current bound, even
   the next generation of experiments might provide some information
   on the Dirac CP phase and the type of the neutrino mass hierarchy.
\end{abstract}

\vspace*{.5cm}

\end{titlepage}

\newpage

\renewcommand{\thefootnote}{\arabic{footnote}}
\setcounter{footnote}{0}

%%%%%%%%%%%%%%%%%%%%%%%%%%%%%%%%%%%%%%%%%%%%%%%%%%%%%%%%%%%%%%%%%%%%%
%                     Introduction                                  %
%%%%%%%%%%%%%%%%%%%%%%%%%%%%%%%%%%%%%%%%%%%%%%%%%%%%%%%%%%%%%%%%%%%%%

\section{Introduction}

Within the last ten years a huge progress has been achieved in
neutrino oscillation physics.  In particular, the results of the
atmospheric neutrino
experiments~\cite{Fukuda:1998mi,SKupdate,Ambrosio:2003yz}
and the K2K accelerator neutrino experiment~\cite{Ahn:2002up} have
demonstrated that atmospheric muon neutrinos oscillate predominately
into tau neutrinos with a mixing angle close to maximal
mixing. Furthermore, solar neutrino
experiments~\cite{Cleveland:1998nv,Ahmed:2003kj}
and the KamLAND reactor neutrino experiment~\cite{Eguchi:2002dm} have
established that the reduced flux of solar electron neutrinos
is consistently understood by the so-called LMA-MSW
solution~\cite{Wolfenstein:1978ue}.
Looking back at these exciting developments, it is tempting to
extrapolate where we could stand in ten years from now with the
experiments being under construction or planned. Certainly, neutrino
physics will turn from the discovery era to the precision age, which
however, will make this field by no means less exciting. The next
major challenge will be the determination of the third, unknown mixing
angle $\theta_{13}$, which at present is only known to be
small~\cite{Apollonio:1999ae,Apollonio:2002gd}. Further important
issues will be the determination of the neutrino mass hierarchy and,
if $\theta_{13}$ turns out to be large enough, the Dirac CP phase.
Three different classes of experiments are under discussion for the
next generation of long-baseline oscillation experiments, which are
able to address at least some of these topics: Conventional beam
experiments, first-generation superbeams, and new reactor experiments
with near and far detectors. In this study, we consider specific
proposals for such experiments, which are under construction or in
active preparation, and could deliver physics results within the next
ten years.
 
An already existing conventional beam experiment is the K2K
experiment~\cite{Ahn:2002up}, which is sending a neutrino beam from
the KEK accelerator to the Super-Kamiokande detector. This experiment
has already confirmed the disappearance of $\nu_\mu$ as predicted by
atmospheric neutrino data, and with more statistics it will slightly
reduce the allowed range of the atmospheric mass splitting $\ldm$. In
this study, we consider in detail the next generation of such
conventional beam experiments, which are the \minos\
experiment~\cite{Ables:1995wq} in US, and the CERN to Gran Sasso
(CNGS) experiments \icarus~\cite{Aprili:2002wx} and
\opera~\cite{Duchesneau:2002yq}. These experiments are currently under
construction and should easily obtain physics results within the next
ten years, including five years of data taking.

Moreover, we consider the subsequent generation of beam experiments, the
so-called superbeam experiments. They use the same technology as
conventional beams with several improvements. 
The most advanced superbeam proposals are the J-PARC to
Super-Kamiokande experiment (\JHFSK)~\cite{Itow:2001ee} in Japan, and
the \NuMI\ off-axis experiment~\cite{Ayres:2002nm}, using a neutrino
beam produced at Fermilab in US.  For these two experiments 
specific Letters of Intent exist and we use the setups discussed in
there. \JHFSK\ and \NuMI\ could deliver important new results towards
the end of the timescale considered in this work.

Recently, there has been a lot of activity to investigate the potential of
new reactor neutrino experiments~\cite{whitepaper}. It has been
realized that the performance of previous experiments, such as
CHOOZ~\cite{Apollonio:1999ae,Apollonio:2002gd} or Palo
Verde~\cite{Boehm:2001ik}, can be significantly improved if a near
detector is used to control systematics and if the statistics is
increased~\cite{Mikaelyan:1999pm,Minakata:2002jv,Huber:2003pm}.
A number of possible sites are discussed, including reactors in
Brasil, China, France, Japan, Russia, Taiwan, and the US (see
\Ref~\cite{whitepaper} for an extensive review). Among the discussed
options are the KASKA project in Japan~\cite{Minakata:2002jv} at the
Kashiwazaki-Kariwa power plant, several power plants in
USA~\cite{Shaevitz:2003ws,Heeger} (\eg, Diablo Canyon in California or
Braidwood in Illinois), and the \DChooz\ project~\cite{doubleChooz}
(\CHOOZII), which is planned at the original CHOOZ
site~\cite{Apollonio:2002gd} in France.

The particular selection of experiments considered in this study is
determined by the requirement that results should be available within
about ten years from now. This either requires that the experiments
are already under construction (such as \minos, \icarus, and \opera), or
that specific proposals (Letters of Intent) including feasibility
studies exist. From the current perspective, the only superbeam
experiments fulfilling this requirement are the \JHFSK\ and \NuMI\
projects. Concerning reactors, we consider in this study the \DChooz\
project~\cite{doubleChooz}, since this proposal has the advantage that
a lot of infrastructure from the first CHOOZ experiment can be
re-used. In particular, the existence of the detector hall drastically
reduces the required amount of civil engineering, which is
considered to be time-critical for a future reactor
experiment. Therefore, it seems rather likely that a medium size
experiment can be built at the CHOOZ site within a few years and
deliver physics results during the timescale considered here. We would
like to stress that other reactor experiments of similar size, such as
the KASKA project in Japan~\cite{Minakata:2002jv}, would lead to
results similar to \DChooz. To fully explore the potential of neutrino
oscillation experiments at nuclear reactors, we furthermore consider
an even larger reactor neutrino experiment (\ReactorII). This could be
especially interesting if a large value of $\theta_{13}$ was
found. \ReactorII\ is the only exception for which we use an abstract
setup, which could, in principle, be built at one of the sites
mentioned above. For example, some projects discussed in the US, such
as Diablo Canyon or Braidwood~\cite{Heeger,shaevitztalk}, are similar
to our \ReactorII\ setup. Such an experiment could be feasible within
a timescale similar to the superbeam experiments, and could provide
results at the end of the period considered in this work. Note that 
in this study, we do not consider oscillation experiments using a 
natural neutrino source, such as solar, atmospheric, or supernova neutrinos.

The outline of the paper is as follows: After a brief description of
the considered experiments in \Sec~\ref{sec:expclasses}, we discuss
the analysis methods and some analytical qualitative features of our
results in \Sec~\ref{sec:ana}. The main results of this study are
given in \Secs~\ref{sec:th23ldm}, \ref{sec:stheta_conv},
\ref{sec:theta13_all}, and \ref{sec:th13delta}. First, in
\Sec~\ref{sec:th23ldm}, we investigate the improvement of the
atmospheric parameters $\theta_{23}$ and $\ldm$ from long-baseline
experiments within ten years. Then we move to the discussion of the
$\stheta$ sensitivity limit if no finite value of $\stheta$ can be
established. We consider in \Sec~\ref{sec:stheta_conv} the
conventional beam experiments \minos, \icarus, and \opera. In
\Sec~\ref{sec:theta13_all}, we discuss the potential of reactor
neutrino experiments to constrain $\stheta$, and we compare the final
$\stheta$ bounds from the conventional beams, \DChooz, \JHFSK, and
\NuMI . In \Sec~\ref{sec:th13delta}, we investigate the assumption
that $\stheta$ is large, and discuss what we could learn from the next
generation of experiments on the Dirac CP phase and the type of the
neutrino mass hierarchy. In this section, the \ReactorII\ setup will
become important. A summary of our results is given in
\Sec~\ref{sec:conclusions}. In \App~\ref{sec:simbeams}, we describe in
detail our simulation of \minos, \icarus, and \opera . Furthermore, in
\App~\ref{app:reactor}, technical details of the reactor experiment
analysis are given. Eventually, we present a thorough discussion of
our definition of the $\stheta$ limit in \App~\ref{app:stheta}.

%%%%%%%%%%%%%%%%%%%%%%%%%%%%%%%%%%%%%%%%%%%%%%%%%%%%%%%%%%%%%%%%%%%%%%%%%%%
%%%%%%%                     The Experiments                          %%%%%%
%%%%%%%%%%%%%%%%%%%%%%%%%%%%%%%%%%%%%%%%%%%%%%%%%%%%%%%%%%%%%%%%%%%%%%%%%%%

\section{Description of the considered experiments}
\label{sec:expclasses}

In this section, we discuss in detail the individual experiments
considered in this work. The main characteristics of the used setups
are summarized in \Tab~\ref{tab:reps}.

\subsection{Conventional beam experiments}

Conventional beam experiments use an accelerator for neutrino
production: A proton beam hits a target and produces a pion beam (with
a contribution of kaons). The resulting pions mainly decay into muon
neutrinos with some electron neutrino contamination. The far detector
is usually located in the center of the beam. The primary goal of
these beams is the improvement of the precision of the atmospheric
oscillation parameters. In addition, an improvement of the CHOOZ limit
for $\stheta$ is expected. For more details, see
\Ref~\cite{Ables:1995wq} for the \minos\ experiment and
\Refs~\cite{Aprili:2002wx,Duchesneau:2002yq} for the CNGS
experiments. In addition, we describe our simulation in more detail in
\App~\ref{sec:simbeams}.

The neutrino beam for the \minos\ experiment is produced at Fermilab.
Protons with an energy of about $120\,\mathrm{GeV}$ hit a graphite
target with an intended exposure of $3.7 \cdot 10^{20}$ protons on
target (pot) per year. A two-horn focusing system allows to direct the
pions towards the Soudan mine where the magnetized iron far detector
is located, which results in a baseline of $735\,\mathrm{km}$. The
flavor content of the beam is, because of the decay characteristics of
the pions, almost only $\nu_\mu$ with a contamination of approximately
1\% $\nu_e$. The mean neutrino energy is at $\langle
E_\nu \rangle \sim 3\,\mathrm{GeV}$, which is small compared to the
$\tau$-production threshold.  The main purpose is to observe $\nu_\mu
\rightarrow \nu_\mu$ disappearance with high statistics, and thus to
determine the ``atmospheric'' oscillation parameters. In addition, the 
$\nu_\mu \rightarrow \nu_e$ appearance channel will provide some
information on $\stheta$.

The CNGS beam is produced at CERN and directed towards the Gran Sasso
Laboratory, where the \icarus\ and \opera\ detectors are located at a
baseline of $732\,\mathrm{km}$. The primary protons are accelerated in
the SPS to $400\,\mathrm{GeV}$, and the luminosity is planned to be
$4.5 \cdot 10^{19}\,\mathrm{pot}\,\mathrm{y}^{-1}$.  Again the beam
mainly contains $\nu_\mu$ with a small contamination of $\nu_e$ at the
level of 1\%. The main difference to the \NuMI\ beam is the higher neutrino
energy. The mean energy is $17\,\mathrm{GeV}$, well above the
$\tau$-production threshold. Therefore, the CNGS experiments will be
able to study the $\nu_\tau$-appearance in the
$\nu_\mu\rightarrow\nu_\tau$ channel.  Two far detectors with very
different technologies designed for $\nu_\tau$ detection will be used
for the CNGS experiment. The \opera\ detector is an emulsion cloud
chamber, whereas \icarus\ is based on a liquid Argon TPC.
In addition to the $\nu_\tau$ detection, it is possible to identify
electrons and muons in the \opera\ and \icarus\ detectors. This in addition
allows to study the $\nu_\mu\to\nu_e$ appearance channel providing the
main information on $\stheta$, and the $\nu_\mu$ disappearance
channel, which contributes significantly to the determination of the
atmospheric oscillation parameters.

\subsection{The first-generation superbeams \JHFSK\ and \NuMI\ }

Superbeams are based upon the technology of conventional beam
experiments with some technical improvements. All superbeams use a
near detector for a better control of the systematics and are aiming
for higher target powers than the conventional beam experiments. In
addition, the detectors are better optimized for the considered
purpose. Since the primary goal of superbeams is the $\stheta$
sensitivity, the $\nu_\mu \rightarrow \nu_e$ appearance channel is
expected to provide the most interesting results. In order to reduce
the irreducible fraction of $\nu_e$ from the meson decays (which is
also called ``background'') and the unwanted high-energy tail in the
neutrino energy spectrum, one uses the {\em
off-axis}-technology~\cite{offaxis} to produce a narrow-band beam,
\ie, a neutrino beam with a sharply peaking energy spectrum. For this
technology, the far detector is situated slightly off the beam
axis. The simulation of the superbeams is performed as described in
\Ref~\cite{Huber:2002rs}; here we give only a short summary.

The J-PARC to Super-Kamiokande superbeam, which we further on call
\JHFSK,\footnote{The \JHFSK\ setup considered in this work is
the same as the setup labeled {\sf JHF-SK} in previous
publications~\cite{Huber:2002mx,Huber:2002rs,Huber:2003pm}.} is
supposed to have a target power of $0.77 \, \mathrm{MW}$ with $10^{21}
\, \mathrm{pot}$ per year~\cite{Itow:2001ee}. It uses the
Super-Kamiokande detector, a water Cherenkov detector with a fiducial
mass of $22.5 \, \mathrm{kt}$ at a baseline of $L=295 \, \mathrm{km}$
and an off-axis angle of $2^\circ$. The Super-Kamiokande detector has
excellent electron-muon separation and neutral current rejection
capabilities. Since the mean neutrino energy is $0.76 \,
\mathrm{GeV}$, quasi-elastic scattering is the dominant detection
process.

\begin{table}[t]
\begin{center}
\begin{tabular}{|lrrrlrr|}
\hline
Label & $L$ & $\langle E_\nu \rangle$ & $P_{\mathrm{Source}}$ & 
Detector technology & $m_{\mathrm{Det}}$ & $t_{\mathrm{run}}$ \\
\hline
\multicolumn{7}{|l|}{\bf{Conventional beam experiments:}} \\
\minos\ & $735 \, \mathrm{km}$ & $3 \,\mathrm{GeV}$ & 
$3.7 \cdot 10^{20} \,\mathrm{pot/y}$ & 
Magn. iron calorim. &  $5.4\,\mathrm{kt}$ & $5 \, \mathrm{yr}$ \\
\icarus\ & $732\,\mathrm{km}$ &  $17\,\mathrm{GeV}$  & 
$4.5 \cdot 10^{19}\,\mathrm{pot/y}$ & 
Liquid Argon TPC & $2.35\,\mathrm{kt}$ & $5 \, \mathrm{yr}$\\
\opera\ & $732\,\mathrm{km}$ &  $17\,\mathrm{GeV}$ & 
$4.5 \cdot 10^{19}\,\mathrm{pot/y}$ &  
Emul. cloud chamb. &  $1.65\,\mathrm{kt}$ & $5 \, \mathrm{yr}$\\[0.1cm]
\multicolumn{7}{|l|}{\bf{Superbeams:}} \\
\JHFSK\ & $295  \, \mathrm{km}$ & $0.76 \, \mathrm{GeV}$ & 
$1.0 \cdot 10^{21} \, \mathrm{pot/y}$  & 
Water Cherenkov & $22.5 \, \mathrm{kt}$ & $5 \, \mathrm{yr}$ \\
\NuMI\ & $812 \, \mathrm{km}$ & $2.22 \, \mathrm{GeV}$ & 
$4.0 \cdot 10^{20} \,\mathrm{pot/y}$ & 
Low-Z-calorimeter & $50 \, \mathrm{kt}$ & $5 \, \mathrm{yr}$ \\[0.1cm]
\multicolumn{7}{|l|}{\bf{Reactor experiments:}} \\
\CHOOZII\ & $1.05 \, \mathrm{km}$ & $\sim 4 \, \mathrm{MeV}$ & $2
\times 4.25 \, \mathrm{GW}$ & 
Liquid Scintillator & $11.3 \, \mathrm{t}$ & $3 \, \mathrm{yr}$ \\
\ReactorII\ & $1.70 \, \mathrm{km}$ & $\sim 4 \, \mathrm{MeV}$ & 
$ 8\,\mathrm{GW}$ & 
Liquid Scintillator & $200\,\mathrm{t}$ & $5\,\mathrm{yr}$ \\ 
\hline
\end{tabular}
\end{center}
\mycaption{\label{tab:reps} The different classes of experiments and
   the considered setups. The table shows the label of the experiment,
   the baseline $L$, the mean neutrino energy $\langle E_\nu \rangle$,
   the source power $P_{\mathrm{Source}}$ (for beams: in protons on
   target per year, for reactors: in gigawatts of thermal reactor
   power), the detector technology, the fiducial detector mass
   $m_{\mathrm{Det}}$, and the running time $t_{\mathrm{run}}$. Note
   that most results are, to a first approximation, a function of the
   product of running time, detector mass, and source power.}
\end{table}

For the NuMI off-axis experiment~\cite{Ayres:2002nm}, which we further
on call \NuMI , a low-Z-calorimeter with a fiducial mass of $50 \,
\mathrm{kt}$ is planned~\cite{NuMI}. Because of the higher average
neutrino energy of about $2.2 \, \mathrm{GeV}$, deep inelastic
scattering is the dominant detection process. Thus, the hadronic
fraction of the energy deposition is larger at these energies, which
makes the low-Z-calorimeter the more efficient detector technology.
For the baseline and off-axis angle, many configurations are under
discussion. As it has been demonstrated in
\Refs~\cite{Barger:2002xk,Huber:2002rs,Minakata:2003ca}, a \NuMI\
baseline significantly longer than $712 \, \mathrm{km}$ increases the
overall physics potential because of the larger contribution of matter
effects. In this study, we use a baseline of $812 \, \mathrm{km}$ and
an off-axis angle of $0.72^\circ$, which corresponds to a location
close to the proposed Ash River site, and to the longest possible
baseline within the United States.
%For a longer baseline, one had to go to Canada (Ontario). 
The beam is supposed to have a target power of about $0.43 \,
\mathrm{MW}$ with $4.0 \cdot 10^{20} \, \mathrm{pot}$ per year.

\subsection{The reactor experiments \DChooz\ and \ReactorII\ }

The key idea of the new proposed reactor experiments is the use of a
near detector at a distance of few hundred meters away from the
reactor core. If near and far detectors are built as identical as
possible, systematic uncertainties related to the neutrino flux will
cancel. In addition, detectors considerably larger than the CHOOZ
detector are anticipated, which has, for example, been demonstrated to
be feasible by KamLAND~\cite{Eguchi:2002dm}. Except from these
improvements, such a reactor experiment would be very similar to
previous experiments, such as CHOOZ~\cite{Apollonio:2002gd} or Palo
Verde~\cite{Boehm:2001ik}. The basic principle is the detection of
antineutrinos by the inverse $\beta$-decay process, which are produced by $\beta$-decay in a nuclear fission reactor. For details of our
simulation of reactor neutrino experiments, see
\Ref~\cite{Huber:2003pm} and \App~\ref{app:reactor}.

For the \DChooz\ experiment, we assume a total number of $60 \, 000$
un-oscillated events in the far detector~\cite{doubleChooz}, which
corresponds (for 100\% detection efficiency) to the integrated
luminosity of $288 \, \mathrm{t \cdot GW \cdot yr}$, compared to the original
CHOOZ experiment with $12.25 \, \mathrm{t \cdot GW \cdot yr}$ leading to
about $2\, 500$ un-oscillated events~\cite{Apollonio:1999ae}. The
integrated luminosity is given as the product of thermal reactor
power, running time, and detector mass. Note that, at least for a
background-free measurement, one can scale the individual factors
such that their product remains constant. The possibility to re-use
the cavity of the original CHOOZ experiment is a striking feature of
the \DChooz\ proposal, although it confines the far detector to a
baseline of $1.05\,\mathrm{km}$, which is slightly too short for the
current best-fit value $\ldm \simeq 2\cdot 10^{-3}\,\mathrm{eV}^2$.

If a positive signal for $\stheta$ is found soon, \ie, $\stheta$ turns
out to be large, it will be the primary objective to push the
knowledge on $\stheta$ and $\deltacp$ with the next generations of
experiments. From the initial measurements of superbeams, $\stheta$ and
$\deltacp$ will be highly correlated (see \Sec~\ref{sec:th13delta}).
In order to disentangle these parameters, some complementary
information is needed. For this purpose, one can either use extensive
antineutrino running at a beam experiment, or use an additional large
reactor experiment to measure $\stheta$
precisely~\cite{Huber:2003pm,Minakata:2003wq}. Because the
antineutrino cross sections are much smaller than the neutrino cross
sections, a superbeam experiment would have to run about three times
longer in the antineutrino mode than in the neutrino mode in order to
obtain comparable statistical information. Thus, a superbeam could not
supply the necessary information within the anticipated timescale. We
therefore suggest the large reactor experiment \ReactorII\ from
\Ref~\cite{Huber:2003pm} at the optimal baseline of $L=1.7 \,
\mathrm{km}$ in order to demonstrate the combined potential of all
experiments. It has $636 \, 200$ un-oscillated events, which
corresponds to an integrated luminosity of $8 \, 000 \, \mathrm{t \cdot GW
\cdot yr}$. Such a reactor experiment could, for example, be built at the
Diablo Canyon or Braidwood power plants~\cite{Heeger,shaevitztalk}.

%%%%%%%%%%%%%%%%%%%%%%%%%%%%%%%%%%%%%%%%%%%%%%%%%%%%%%%%%%%%%%%%%%%%%%%%%%%%%
%%%%%%%%%%                     Description of Calculations                   % 
%%%%%%%%%%%%%%%%%%%%%%%%%%%%%%%%%%%%%%%%%%%%%%%%%%%%%%%%%%%%%%%%%%%%%%%%%%%%%

\section{Qualitative discussion and analysis methods}
\label{sec:ana}

In general, our calculations are done in the three flavor framework,
where we use the standard parameterization $U$ of the leptonic mixing
matrix described by three mixing angles and one CP
phase~\cite{PDG}. Our results are based on a full numerical simulation
of the exact transition probabilities, and we also include Earth
matter effects~\cite{Wolfenstein:1978ue} because of the
long baselines used for the \NuMI\ beam. We take into account matter
density uncertainties by imposing an error of $5\%$ on the average
matter density~\cite{Geller:2001ix}.
The probabilities are convoluted with the neutrino fluxes, detection
cross sections, energy resolutions, and experimental efficiencies to
calculate the event rates, which are the basis of the full statistical
$\chi^2$-analysis.  We use all the information available, \ie, the
appearance and disappearance channels, as well as the energy
information. 
The simulation methods are described in the Appendices of
\Ref~\cite{Huber:2002mx}; for details of the conventional beam
experiments, see also \App~\ref{sec:simbeams}, for the superbeam
experiments \Ref~\cite{Huber:2002rs}, and for the the reactor
experiments \Ref~\cite{Huber:2003pm} and \App~\ref{app:reactor}.  All
of the calculations are performed with the GLoBES
software~\cite{Globes}.

In order to obtain a qualitative analytical understanding of the
effects, it is sufficient to use simplified expressions for the
transition probabilities, which are obtained by expanding the
probabilities in vacuum simultaneously in the mass hierarchy parameter
$\alpha \equiv \Delta m_{21}^2 / \Delta m_{31}^2$ and the small mixing
angle $\sin2\theta_{13}$.  The expression for the $\nu_\mu \rightarrow
\nu_e$ appearance probability up to second order in $\alpha$ and
$\sin2\theta_{13}$ is given
by~\cite{Freund:2001ui,Akhmedov:2004ny}
\begin{eqnarray}
P(\nu_\mu \rightarrow \nu_e) & \simeq & \sin^2 2\theta_{13} \, \sin^2 \theta_{23}
\sin^2 {\Delta} \nonumber \\
& \mp &  \alpha\; \sin 2\theta_{13} \, \sin\deltacp  \, 
\sin 2\theta_{12} \sin 2\theta_{23}
\, \Delta \sin^2{\Delta} \nonumber \\
&+&  \alpha\; \sin 2\theta_{13}  \, \cos\deltacp \, 
\sin 2\theta_{12} \sin 2\theta_{23}
\, \Delta \cos {\Delta} \sin {\Delta} \nonumber  \\
&+& \alpha^2 \, \cos^2 \theta_{23} \sin^2 2\theta_{12} \, \Delta^2
\label{equ:beam}
\end{eqnarray} 
with $\Delta \equiv \Delta m^2_{31}L / (4E_\nu)$. The sign of the
second term is negative for neutrinos and positive for antineutrinos.
The relative weight of each of the individual terms in \equ{beam} is
determined by the values of $\alpha$ and $\sin2\theta_{13}$,
which means that the superbeam performance is highly affected by the
true values $\sdm$ and $\ldm$ given by nature. 
Reactor experiments can be described by the corresponding expansion of
the disappearance probability up to second order in $\sin 2
\theta_{13}$ and $\alpha$~\cite{Minakata:2002jv,Huber:2003pm,Akhmedov:2004ny}
\begin{equation}
1 - P_{\bar{e} \bar{e}} \quad \simeq \quad \sin^2 2 \theta_{13} \,
\sin^2 \Delta + \alpha^2 \, \Delta^2 \, \cos^4 \theta_{13}
\, \sin^2 2 \theta_{12} . \label{equ:reactor}
\end{equation}
The second term on the right-hand side of this equation is for
$\stheta \gtrsim 10^{-3}$ and close to the first atmospheric
oscillation maximum relatively small compared to the first one, and
can therefore be neglected in the relevant parameter space region. In
principle, there are also terms of the order $\alpha \, \sin^2 2
\theta_{13}$ and higher orders in \equ{reactor}. Though some of these
terms could be of the order of the $\alpha^2$-term for large values of
$\stheta$, they are, close to the atmospheric oscillation maximum,
always suppressed compared to the $\stheta$-term by at least one order
of $\alpha$. Thus, the $\stheta$-term carries the main information.

From \equ{reactor}, it is obvious that a reactor experiment cannot
access $\theta_{23}$, the mass hierarchy, or $\deltacp$.  In
addition, the measurements of $\ldm$ would only be possible for large
values of $\stheta$~\cite{Huber:2003pm}. These parameters can be only 
measured by the $\nu_\mu \to \nu_\mu$, $\nu_\mu \to \nu_e$, and $\nu_\mu \to
\nu_\tau$ channels in beam experiments. However, comparing
\eqs~(\ref{equ:beam}) and~(\ref{equ:reactor}), one can easily see that reactor
experiments should allow a ``clean'' and degenerate-free measurement
of $\stheta$~\cite{Minakata:2002jv}. In contrast, the determination of
$\stheta$ using the appearance channel in \equ{beam} is strongly
affected by the more complicated parameter dependence of the
oscillation probability, which leads to multi-parameter
correlations~\cite{Huber:2002mx} and to the $(\delta,
\theta_{13})$~\cite{Burguet-Castell:2001ez}, $\mathrm{sgn}(\Delta
m_{31}^2)$~\cite{Minakata:2001qm}, and
$(\theta_{23},\pi/2-\theta_{23})$~\cite{Fogli:1996pv} degeneracies,
\ie, an overall ``eight-fold'' degeneracy~\cite{Barger:2001yr}. In the
analysis, we take into account all of these degeneracies. Note
however, that the $(\theta_{23},\pi/2-\theta_{23})$ degeneracy is not
present, since we always adopt for the true value of $\theta_{23}$ the
current atmospheric best-fit value $\theta_{23}=\pi/4$. The proper
treatment of correlations and degeneracies is of particular importance
for the calculation of a sensitivity limit on $\stheta$. This issue is
discussed in detail in \App~\ref{app:stheta}, where we give also a
precise definition of the $\stheta$ sensitivity limit. In some cases
we compare the actual $\stheta$ sensitivity limit to the so-called
$(\sin^2 2 \theta_{13})_{\mathrm{eff}}$ sensitivity limit, which
includes only statistical and systematical errors (but no correlations
and degeneracies). This limit corresponds roughly to the potential of
a given experiment to observe a positive signal, which is
``parameterized'' by some (unphysical) mixing parameter $(\sin^2 2
\theta_{13})_{\mathrm{eff}}$ (see also \App~\ref{app:stheta} for a
precise definition).

If not otherwise stated, we use in the following for the ``solar'' and
``atmospheric'' parameters the current best-fit values with their
$3\,\sigma$-allowed ranges:
\begin{eqnarray}
\label{eq:bfp}
|\Delta m_{31}^2| = 2.0_{-0.9}^{+1.2} \cdot 10^{-3}\,\mathrm{eV}^2,  &\quad&
\sin^22\theta_{23} = 1_{-0.15}^{+0}, \nonumber\\
\Delta m_{21}^2 = 7.0_{-1.6}^{+2.5} \cdot 10^{-5}\,\mathrm{eV}^2, &\quad&
\sin^22\theta_{12} = 0.8_{-0.1}^{+0.15}.
\end{eqnarray} 
The numbers are taken from
\Refs~\cite{Fogli:2003th,Maltoni:2003da}, which include
the latest SNO salt solar neutrino data~\cite{Ahmed:2003kj} and the
results of the re-analysis of the Super-Kamiokande atmospheric
neutrino data~\cite{SKupdate}.  The interesting dependencies on the
true parameter values are usually shown within the $3 \sigma$-allowed
ranges. For the upper bound on $\stheta$ at 90\% CL ($3\sigma$) we use
\begin{equation}
\label{equ:th13_bound}
\sin^22\theta_{13} \le 0.14\, (0.25) \,,
\end{equation}
obtained from the CHOOZ data~\cite{Apollonio:1999ae} combined with
global solar neutrino and KamLAND data at the best fit value $\ldm =
2\cdot 10^{-3}\,\mathrm{eV}^2$~\cite{Maltoni:2003da}. In order to take
into account relevant information from experiments not considered
explicitly, we impose external input given by the $1\sigma$ error on
the respective parameters. This is mainly relevant for the
``solar parameters'', where we assume that the ongoing KamLAND
experiment will improve the errors down to a level of about $10\%$ on
each $\dm{21}$ and
$\sin2\theta_{12}$~\cite{Gonzalez-Garcia:2001zy}.  For
the ``atmospheric parameters'' we assume as external input roughly the
current error of $20\%$ for $|\dm{31}|$ and $5\%$ for
$\sin^22\theta_{23}$, which however, becomes irrelevant after about
one year of data taking of the conventional beams, since then these
parameters (especially $|\ldm|$) will be determined to a better
precision from the experiments themselves. Furthermore, we assume a
precision of $5\%$ for $|\dm{31}|$ for the separate analysis of the
reactor experiments, since the conventional beams should supply results
until then.  However, it can be shown that the results would only
marginally change for an error of $20\%$ for $|\dm{31}|$.

In general our results presented in the following depend on the assumed
true values of the oscillation parameters. In particular they show a
strong dependence on the true value of $\dm{31}$, and therefore this
dependence will be depicted in figures where appropriate. The
$\theta_{13}$ sensitivity limit obtained from $P_{e\mu}$ moreover also
depends strongly on the true value of $\dm{21}$ (see
\Fig~\ref{fig:solardep} below). In principle also the variation of
$\theta_{12}$ plays a role. However, $P_{e\mu}$ depends only on the
product of $\alpha\cdot \sin2\theta_{12}$ up to second order in
$\alpha$ as shown in \eq~(\ref{equ:beam}). Therefore a variation of
the true value of $\theta_{12}$ is equivalent to a rescaling of the
true value of $\dm{21}$. The variation of the true value of
$\theta_{23}$ within the range given in \eq~(\ref{eq:bfp}) produces
only slight changes in the results. In particular, those changes are
much smaller than the ones caused by the variation of $\dm{31}$. Thus,
in order to keep the presentation of our results concise, we do not
explicitly discuss the dependence of the results on $\theta_{23}$, and
we always adopt the current best fit value $\theta_{23}= \pi/4$ for
the true value.

%%%%%%%%%%%%%%%%%%%%%%%%%%%%%%%%%%%%%%%%%%%%%%%%%%%%%%%%%%%%%%%%%%%%%
%%%%%%%%%         RESULTS   ATMOS                             %%%%%%% 
%%%%%%%%%%%%%%%%%%%%%%%%%%%%%%%%%%%%%%%%%%%%%%%%%%%%%%%%%%%%%%%%%%%%%

\section{The measurements of $\boldsymbol{\ldm}$ and 
$\boldsymbol{\theta_{23}}$}
\label{sec:th23ldm}

In this section, we investigate the ability of the conventional beam
experiments and superbeams to measure the leading atmospheric parameters
$\ldm$ and $\theta_{23}$. We do not include the reactor experiments in
this discussion, since they are rather insensitive to $\ldm$, and
cannot access $\theta_{23}$ at all. The measurement of these
parameters is dominated by the $\nu_\mu\to\nu_\mu$ disappearance
channel in the beam experiments.

\begin{figure}[t!]
   \centering \includegraphics[width=16cm]{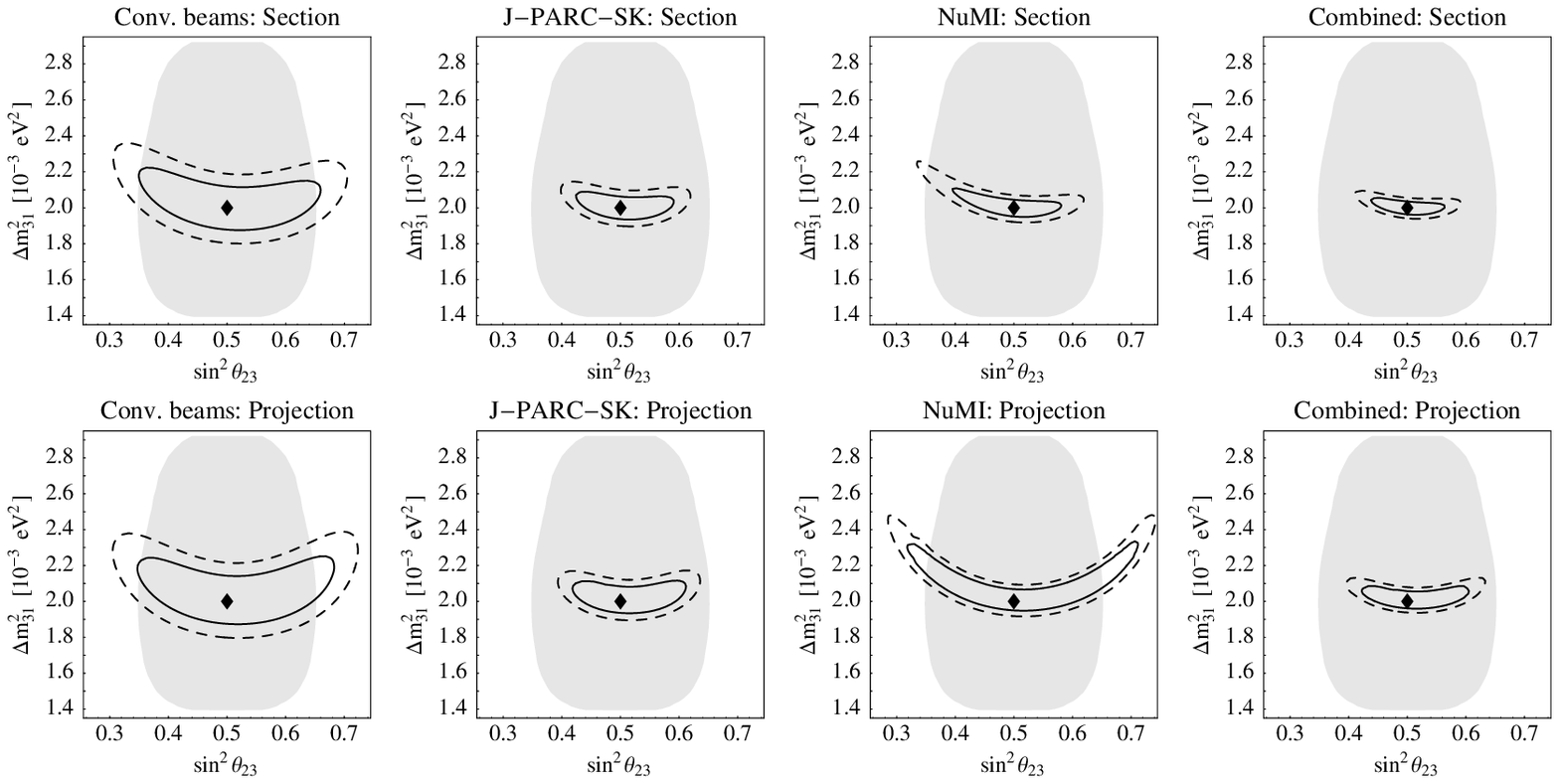}
   \mycaption{\label{fig:dm31th23} The $90 \%$ CL (solid curves) and
   $3 \sigma$ (dashed curves) allowed regions (2 d.o.f.) in the
   $\sin^2 \theta_{23}$-$\ldm$-plane for the combined conventional
   beams (\minos, \icarus, \opera), \JHFSK, \NuMI, and all beam
   experiments combined. For the true values of the oscillation
   parameters, we choose the current best-fit values from
   \eq~(\ref{eq:bfp}), a normal mass hierarchy, $\stheta=0.1$ and
   $\deltacp=0$. The upper row shows a section of the fit manifold (with
   the un-displayed oscillation parameters fixed at their true values),
   and the lower row shows the projection onto the $\sin^2
   \theta_{23}$-$\ldm$-plane as the final result. The shaded
   regions correspond to the 90\% CL allowed region from current
   atmospheric neutrino data~\protect\cite{SKupdate}.} 
\end{figure}

In \figu{dm31th23}, we compare the predicted allowed regions for
$\ldm$ and $\sin^2\theta_{23}$ from the combined conventional beams
(\minos, \icarus, \opera), \JHFSK, \NuMI, and all beam
experiments combined to the current allowed region from
Super-Kamiokande atmospheric neutrino data. We show the fit-manifold
section in the $\sin^2 \theta_{23}$-$\ldm$-plane (upper row), as well
as the projection onto this plane (lower row). For a section, all
oscillation parameters which are not shown are fixed at their true
values, whereas for a projection the $\chi^2$-function is minimized
over these parameters. Therefore, the projection corresponds to the
final result, since it includes the fact that the other fit parameters
are not exactly known. In general, the $\chi^2$-value becomes smaller
by the minimization over the not shown fit parameters, which means
that the allowed regions become larger.
In \figu{dm31th23} the $\mathrm{sgn}(\ldm)$-degeneracy is not
included, since it usually does not produce large effects in the
disappearance channels. In addition, we use the true values
$\stheta=0.1$ and $\deltacp=0^\circ$ in \figu{dm31th23}. Although the
fit-manifold sections shown in the upper row of \figu{dm31th23}
depend to some extent on this choice, the effect for the final
results of the disappearance channels is very small, \ie, the lower
row of \figu{dm31th23} is hardly changed for $\stheta=0$.

The first thing to learn from \figu{dm31th23} is that the precision on
$\ldm$ will drastically improve during the next ten years, whereas our
knowledge on $\theta_{23}$ will be increased rather modestly. The
combination of all the beam experiments will improve the current precision
from the Super-Kamiokande atmospheric neutrino data~\cite{SKupdate} on
$\sin^2\theta_{23}$ roughly by a factor of two, while the precision on
$\ldm$ will be improved by an order of magnitude. Neither the three
conventional beams combined nor \NuMI\ will obtain a precision on
$\theta_{23}$ better than current Super-Kamiokande data, only \JHFSK\
might improve the precision slightly. We note however, that the
$\theta_{23}$ accuracy of the long-baseline experiments strongly
depends on the true value of $\ldm$, and it will be improved if $\ldm$
turns out to be larger than the current best-fit point.

In most cases, the correlations with the un-displayed
oscillation parameters do not cause significant differences between
the sections and projections in the upper and lower rows of
\figu{dm31th23}. Only for \NuMI , the projection is affected by
the multi-parameter correlation with $\stheta$ and $\deltacp$. Since
we do not assume additional knowledge about $\stheta$ for the
individual experiments other than from their own appearance channels,
the appearance channels can indirectly affect the $\ldm$ or
$\theta_{23}$ measurement results. This can be understood in terms of
the disappearance probability, which to leading order is given
by~\cite{Freund:2001ui,Akhmedov:2004ny}
\begin{equation} \label{equ:pmumu}
P_{\mu \mu} = 1 - 
\sin^2 2 \theta_{23} \sin^2 \frac{\Delta m_{31}^2 L}{4 E} 
+ \hdots
\end{equation}
where the dots refer to higher order terms in $\alpha=\sdm/\ldm$ and
$\theta_{13}$, as well as products of these. Thus, the
$\stheta$-precision, which comes from the appearance channels, is
necessary to constrain the amplitude of the higher order terms in this
equation which are proportional to $\theta_{13}$. Since, however,
$\theta_{13}$ is strongly correlated with $\deltacp$ in the appearance
channels, this two-parameter correlation can lead to multi-parameter
correlations with $\theta_{23}$ or $\ldm$ in the disappearance channel
through the higher order terms in \equ{pmumu}. This explains the small
differences between the section and projection plots in
\figu{dm31th23}.
In addition, the measurement of $\stheta$ at \NuMI\ is affected by
matter effects, and hence, is somewhat different for the opposite sign
of $\ldm$. Therefore, one can also expect a slightly different shape
of the fit manifold for the $\mathrm{sgn}(\ldm)$-degeneracy. Note that
the initial asymmetry between $\sin^2 \theta_{23} < 0.5$ and $\sin^2
\theta_{23}>0.5$ for \NuMI\ is caused by its large matter effects.

Eventually, one obtains the precision of the individual parameter
$\ldm$ or $\theta_{23}$ as projection of the lower row plots in
\figu{dm31th23} (for one degree of freedom) onto the respective axis.
In \Tab~\ref{tab:resdmth23} we show our prediction for the $3
\sigma$-allowed ranges of the atmospheric oscillation parameters from
the conventional beam experiments and first generation superbeam
experiments for one degree of freedom.

\begin{table}[t!]
\begin{center}
\begin{tabular}{|l|ccc|}
\hline
Experiment/Combination & $|\ldm|$ & $\theta_{23}$ & $\sin^2 \theta_{23}$ \\
\hline
&&&\\[-0.3cm]
\minos\ + \opera\ + \icarus\ & $2^{+0.34}_{-0.18} \cdot 10^{-3} \, \mathrm{eV}^2$ & $(\pi/4)^{+0.22}_{-0.19}$ & $0.5^{+0.21}_{-0.18}$ \\
&&&\\[-0.3cm]
\JHFSK & $2^{+0.15}_{-0.09} \cdot 10^{-3} \, \mathrm{eV}^2$ & $(\pi/4)^{+0.13}_{-0.10}$ & $0.5^{+0.13}_{-0.10}$ \\
&&&\\[-0.3cm]
\NuMI & $2^{+0.43}_{-0.07} \cdot 10^{-3} \, \mathrm{eV}^2$ & $(\pi/4)^{+0.24}_{-0.21}$ & $0.5^{+0.23}_{-0.20}$ \\
&&&\\[-0.3cm]
All beam experiments combined & $2^{+0.12}_{-0.06} \cdot 10^{-3} \, \mathrm{eV}^2$ & $(\pi/4)^{+0.13}_{-0.10}$ & $0.5^{+0.12}_{-0.09}$ \\
\hline
\end{tabular}
\end{center}
\mycaption{\label{tab:resdmth23} The expected allowed ranges ($3
\sigma$, 1 d.o.f.) for the atmospheric oscillation parameters.  For
the true values of the oscillation parameters, we choose the current
best-fit values, a normal mass hierarchy, $\stheta=0.1$, and
$\deltacp=0^\circ$. The impact of an inverted mass hierarchy, and
different values for $\stheta$ or $\deltacp$ on these final results is
rather small.}
\end{table}

%%%%%%%%%%%%%%%%%%%%%%%%%%%%%%%%%%%%%%%%%%%%%%%%%%%%%%%%%%%%%%%%%%%%%%%%%%%%%%
%%%%%%%%         RESULTS theta13 conventional beams             %%%%%%%%%%%%%%
%%%%%%%%%%%%%%%%%%%%%%%%%%%%%%%%%%%%%%%%%%%%%%%%%%%%%%%%%%%%%%%%%%%%%%%%%%%%%%

\section{Improved $\boldsymbol{\stheta}$ bounds from
  conventional beams}
\label{sec:stheta_conv}

Let us now come to the crucial next step in neutrino oscillation
physics: the determination of the small mixing angle $\theta_{13}$.
We start this discussion by investigating the potential of the 
conventional beam experiments \minos , \icarus , and \opera\ to
improve the current bound on $\sin^22\theta_{13}$.

In \Tab~\ref{tab:chooz}, we show the signal and background event
rates after one year of nominal operation for each experiment
(computed for $\sin^22\theta_{13}=0.1$ and $\delta=0$). Based on these
numbers, one would expect that \minos\ performs significantly better
than \icarus.  However, \Tab~\ref{tab:chooz} only shows integrated
event rates and does not include the energy dependence of signal
versus background event numbers. In the CNGS beam, the energy
distribution of the intrinsic $\nu_e$-contamination is rather
different from the energy distribution of the signal events. Thus, in
a full analysis including energy information, the impact of the
background is reduced. On the other hand, for the NuMI neutrino beam,
the intrinsic $\nu_e$-contamination has an energy distribution which
is much closer to the one of the signal events. Therefore, the impact
of the background is relatively high. 

\begin{table}[t!]
\begin{center}
\begin{tabular}{|l|rrr|}
\hline
&\minos&\icarus&\opera\\
\hline
Signal&7.1&4.4&1.6\\
Background&21.6&12.2&5.4\\
\hline
S/B&0.33&0.36&0.30\\
\hline
\end{tabular}
\end{center}
  \mycaption{\label{tab:chooz}The number of signal and background
  events after one year of nominal operation of \minos, \icarus , and
  \opera . For the oscillation parameters, we use the current best-fit
  values with $\sin^22\theta_{13}=0.1$, $\deltacp = 0$, and a normal mass
  hierarchy.}
\end{table}

\begin{figure}[t!]
\centering \includegraphics[angle=-90,width=\textwidth]{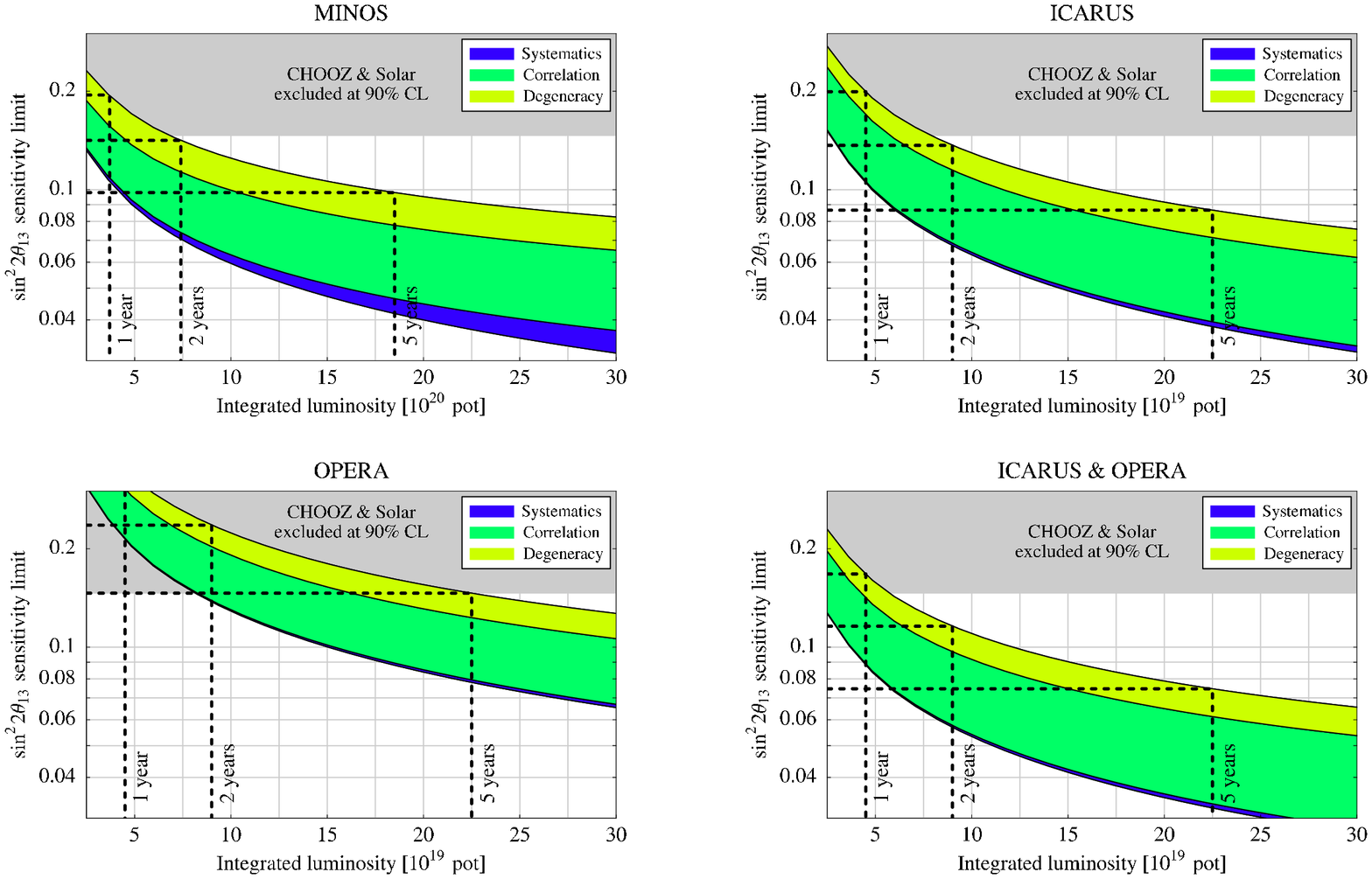}
   \mycaption{\label{fig:lumiscaling}The $\stheta$ sensitivity limit
   as function of the total number of protons on target at the 90\%
   confidence level for \minos , \icarus , \opera , and \icarus\ and
   \opera\ combined (5\% flux uncertainty assumed).  The dashed curves
   refer to the sensitivity limits after one, two, and five years of
   running. The lowest curves are obtained for the statistics limits
   only, whereas the highest curves are obtained after successively
   switching on systematics, correlations, and degeneracies, \ie, they
   correspond to the final sensitivity limits. The gray-shaded area
   shows the current $\sin^22\theta_{13}$ excluded region $\stheta
   \gtrsim 0.14$ at the 90\% CL~\protect\cite{Maltoni:2003da}. For the true
   values of the oscillation parameters we use the current best-fit
   values \eq~(\ref{eq:bfp}) and a normal mass hierarchy.}
\end{figure}

\begin{figure}[t!]
\centering
   \includegraphics[width=10cm]{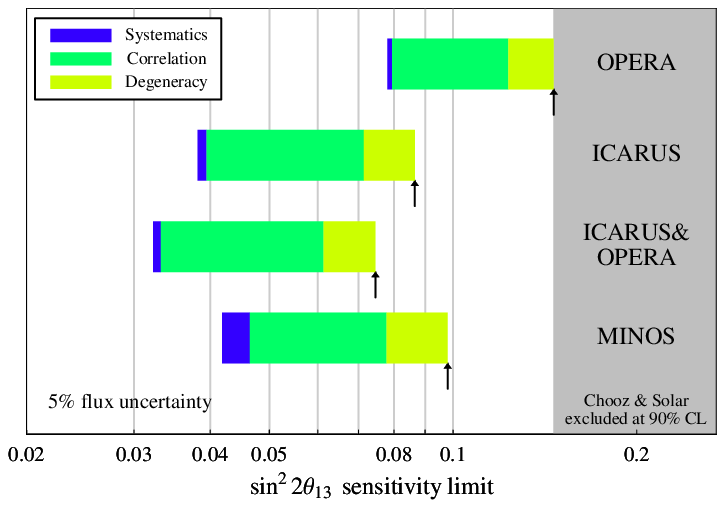} %
   \mycaption{\label{fig:interncomp} The $\stheta$ sensitivity limit
   at the 90\% confidence level after a running time of five years for
   the different experiments. The left edges of the bars are obtained
   for the statistics limits only, whereas the right edges are obtained
   after successively switching on systematics, correlations, and
   degeneracies, \ie, they correspond to the final $\stheta$
   sensitivity limits.  The gray-shaded area shows the current
   $\sin^22\theta_{13}$ excluded region $\stheta \gtrsim 0.14$ at the
   90\% CL~\protect\cite{Maltoni:2003da}. For the true values of the
   oscillation parameters, we use the current best-fit values
   \eq~(\ref{eq:bfp}) and a normal mass hierarchy.}
\end{figure}

An important issue for the $\stheta$ sensitivity limit from the
conventional beams is the finally achieved integrated luminosity,
which might differ significantly from the nominal value due to some
unforeseen experimental circumstances. Therefore, we discuss the
$\stheta$ sensitivity as a function of the integrated number of
protons on target. In \figu{lumiscaling}, the sensitivity limits for
\minos , \icarus , and \opera\ are shown as a function of the
luminosity. Note that since the CNGS experiments will be running
simultaneously, we also show the combined \icarus\ and \opera\
sensitivity limit. In order to compare the achievable limits as a
function of the running time, the dashed lines refer to the results
after one, two, and five years of data taking with the nominal beam
fluxes given in \Refs~\cite{NUMI714,Aprili:2002wx,Komatsu:2002sz}.
The lowest curves are obtained for the statistics limits only, whereas
the highest curves are obtained after successively switching on
systematics, correlations, and degeneracies. Thus, the actual
$\sin^22\theta_{13}$ sensitivity limit in \figu{lumiscaling} is given
by the highest curves. The figure indicates that the CNGS experiments
together can improve the CHOOZ bound after about one and a half years
of running time, and \minos\ after about two years. We note that the
impact of systematics increases for \minos\ with increasing
luminosity, illustrating the typical background problem mentioned
above. In \figu{interncomp}, we eventually summarize the $\stheta$
sensitivity after a total running time of five years for each
experiment, assuming the true value of $\ldm = 2\cdot
10^{-3}\,\mathrm{eV}^2$. One can directly read off this figure that
 the $\stheta$ sensitivity limits of \icarus\ and \minos\ are very similar,
and \icarus\ and \opera\ combined are slightly better than \minos .

Let us briefly compare our results to $\stheta$ sensitivity limit
calculations for \minos, \icarus, and \opera\ existing in the
literature. In the analysis of \Ref~\cite{Migliozzi:2003pw}, the
correlation with $\deltacp$ and the sign($\Delta m_{31}^2$)-degeneracy
are included, and hence these results should be compared with our
final sensitivity limits, although we also include correlations
with respect to all the other oscillation parameters. However, for the
comparison, one has to take into account the different considered running
times for \minos\ (2 years vs. 5 years), as well as the difference in the chosen
true value of $|\Delta m_{31}^2|$ ($3.0\cdot 10^{-3} \,\mathrm{eV}^2$
vs. $2.0\cdot 10^{-3}\,\mathrm{eV}^2$). In the analysis performed in
\Ref~\cite{Barger:2001yx}, the correlation with $\deltacp$ and the
sign($\Delta m_{31}^2$)-degeneracy were not considered, while
correlations with $|\Delta m^2_{31}|$ were taken into
account. Therefore, the results from that study should roughly be 
compared to our $(\sin^22\theta_{13})_\mathrm{eff}$ limits. 
Again one has to take into account different assumptions about the 
running times and the true
value for $\Delta m^2_{21}$ ($5.0\cdot 10^{-5} \,\mathrm{eV}^2$
vs. $7.0\cdot 10^{-5}\,\mathrm{eV}^2$). Finally, in
\App~\ref{app:reproduction} we demonstrate explicitly that our results
are in excellent agreement with the ones obtained by the \minos,
\icarus, and \opera\
collaborations~\cite{NUMI714,Aprili:2002wx,Komatsu:2002sz} if we use
the same assumptions.

\begin{figure}[t!]
\centering \includegraphics[width=10cm]{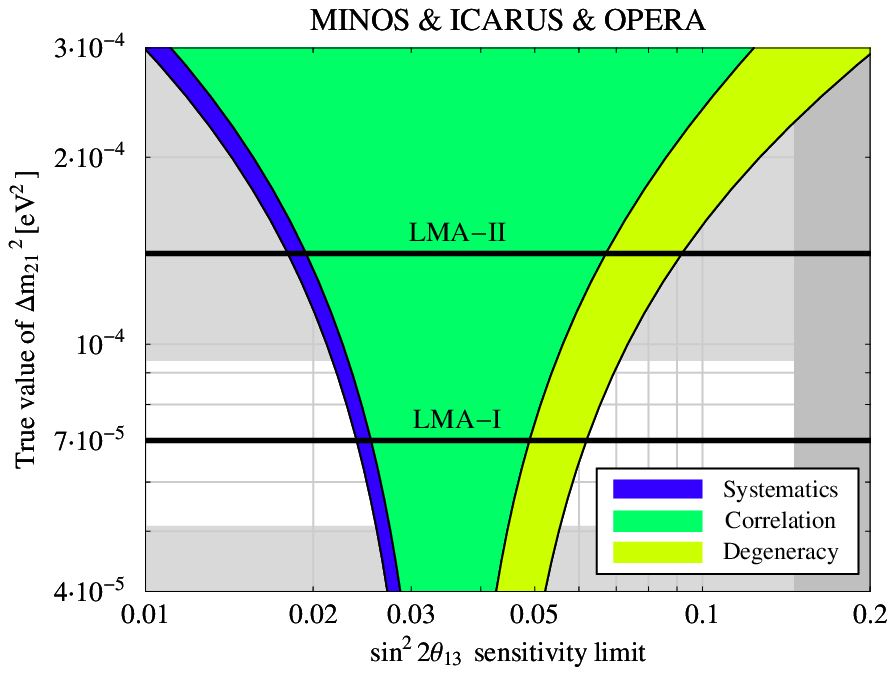}
  \mycaption{\label{fig:solardep} The $\stheta$ sensitivity limit at
  90\% CL for \minos , \icarus , and \opera\ combined as function of
  the true value of $\sdm$ (five years running time). The left curve
  is obtained for the statistics limit only, whereas the right curve is
  obtained after successively switching on systematics, correlations,
  and degeneracies, \ie, it corresponds to the final $\stheta$ sensitivity
  limit.  The dark gray-shaded area shows the current
  $\sin^22\theta_{13}$ excluded region $\stheta \gtrsim 0.14$ at the
  90\% CL, and the light gray-shaded area refers to the LMA-excluded
  region at $3 \sigma$, where the best-fit value is marked by the
  horizontal line~\protect\cite{Maltoni:2003da}. For the true values of the
  un-displayed oscillation parameters we use the current best-fit
  values in \eq~(\ref{eq:bfp}) and a normal mass hierarchy. } 
\end{figure}

A very interesting issue for the conventional beam experiments is the
impact of the true value of $\sdm$ on the $\stheta$ sensitivity. (The
impact of the true value of $\ldm$ is discussed in
\Sec~\ref{sec:theta13_all}.) One can easily see from \equ{beam} that
the effect of $\deltacp$ increases with increasing $\alpha\equiv
\sdm/\ldm$, which determines the amplitude of the second and third
terms in this equation.  Since the main contribution to the
correlation part of the discussed figures comes from the correlation
with $\deltacp$ (with some contribution of the uncertainty of the
solar parameters), a larger $\sdm$ causes a larger correlation
bar. This can clearly by seen from \figu{solardep}, which shows the
combined potential of the conventional beams after five years of
running time (for each experiment) as a function of $\sdm$. In this
figure the right edge of the blue band corresponds to the limit based
only on statistical and systematical errors, \ie, the $(\sin^2 2
\theta_{13})_{\mathrm{eff}}$ sensitivity limit. We find that the
larger $\sdm$ is, the better becomes the systematics-based $(\sin^2 2
\theta_{13})_{\mathrm{eff}}$ sensitivity limit, and the worse becomes the
final sensitivity limit on $\stheta$.  Since the LMA-II region is now
disfavored by the latest solar neutrino and KamLAND data,
\figu{solardep} demonstrates that the conventional beam experiments
can definitively improve the current $\stheta$-bound.  One may expect
an improvement down to $\stheta \lesssim 0.05-0.07$ within the LMA-I allowed
region, where the $\stheta$ sensitivity limit at the current best-fit
value is about $\stheta \le 0.06$.

Since the $\stheta$ sensitivity limit is expected to be $\stheta \le 0.06$ for
\minos , \icarus , and \opera\ combined (with five years running time
for each experiment), a further improvement from the conventional beams
seems to be unlikely. In addition, the systematics limitation, which
can be clearly seen in \figu{lumiscaling}, demonstrates that a further
increase of the luminosity would not lead to significantly better
bounds on $\stheta$. Therefore, one has to proceed to the next generation of
experiments to increase the $\stheta$ sensitivity.  Especially,
the off-axis technology to suppress backgrounds and more optimized
detectors could help to improve the performance.  Amongst other
experiments, we discuss the corresponding superbeams, which are using
these improvements, in the next section.
 
%%%%%%%%%%%%%%%%%%%%%%%%%%%%%%%%%%%%%%%%%%%%%%%%%%%%%%%%%%%%%%%%%%%%%%%%%%%%
%%%%%%%                  RESULTS theta13 ALL                          %%%%%%
%%%%%%%%%%%%%%%%%%%%%%%%%%%%%%%%%%%%%%%%%%%%%%%%%%%%%%%%%%%%%%%%%%%%%%%%%%%%

\section{Further improvement of the $\boldsymbol{\stheta}$ bound}
\label{sec:theta13_all}

After the discussion of the conventional beams in the last section, we
here discuss the final bound on $\stheta$ in ten years from now, if no
finite value will be found (we will in the next section
consider the case of a large $\theta_{13}$). We first discuss in
\Sec~\ref{sec:reactor} the potential of a new reactor neutrino
experiment, whereas we compare in \Sec~\ref{sec:th13_compare} the
$\stheta$ limits from conventional beams, reactor experiments, and
superbeams.

\subsection{Characteristics of reactor neutrino experiments}
\label{sec:reactor}
  
\begin{figure}[t!]
\centering
   \includegraphics[angle=-90,width=\textwidth]{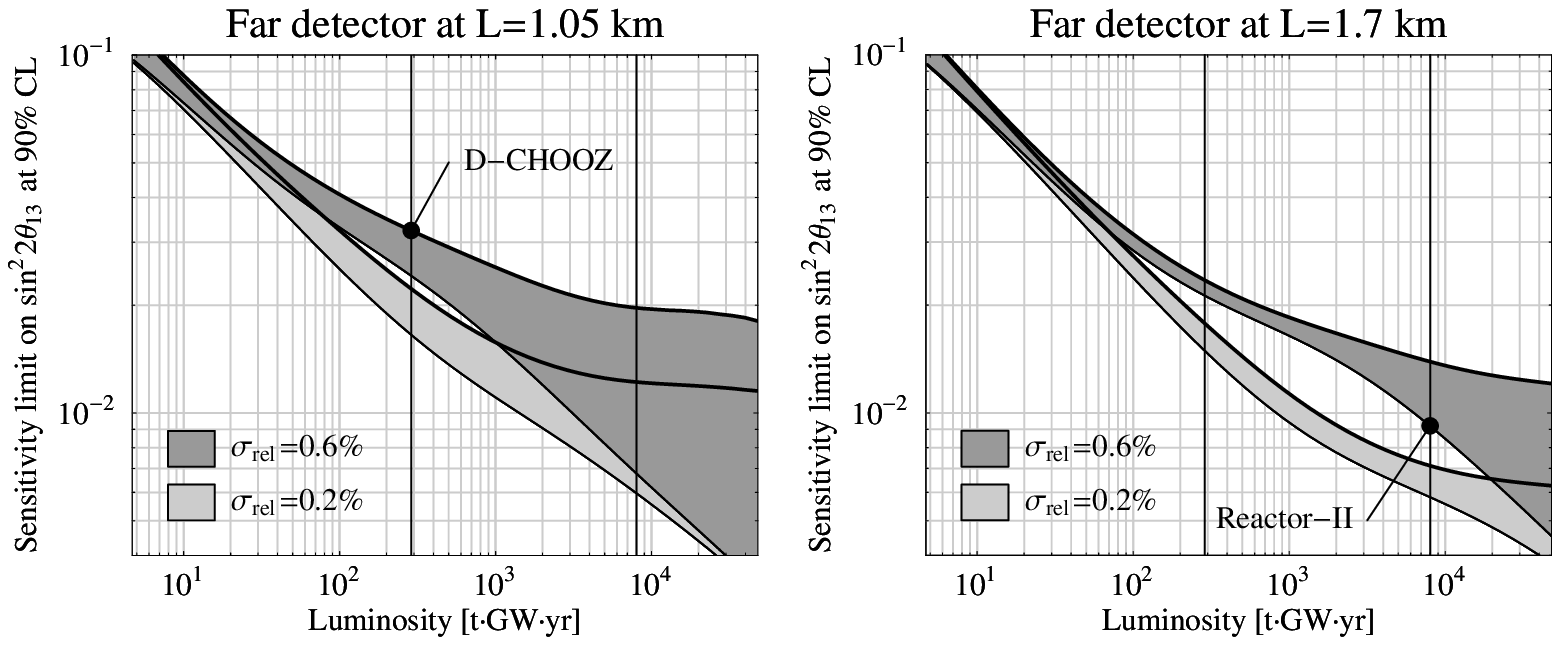}
   \mycaption{\label{fig:lumireact} Luminosity scaling of the
   $\stheta$ sensitivity at the 90\% CL. Here $\ldm = 2 \cdot
   10^{-3}$~eV$^2$ is assumed to be known within 5\%, $L_\mathrm{ND} = 0.15
   \,\mathrm{km}$, and $L_\mathrm{FD} = 1.05 \, (1.7)\, \mathrm{km}$
   in the left (right) panel. The number of events in the near
   detector is fixed to $2.94\cdot 10^6$. We use
   $\sigma_\mathrm{abs} = 2.5\%$ and $\sigma_\mathrm{rel} = 0.2\%
   \,(0.6\%)$ for the light (dark) shaded regions. The upper edge of
   each region is calculated for $\sigma_\mathrm{shape} = 2\%$,
   $\sigma_\mathrm{cal} = 0.5\%$, and backgrounds as given in
   \Tab~\ref{tab:backgrounds} in \App~\ref{app:reactor}. For
   the lower edges, we set $\sigma_\mathrm{shape} = \sigma_\mathrm{cal}
   = 0$ and do not include backgrounds. The dots mark the
   \DChooz\ and \ReactorII\ setups.}
\end{figure}

In \Fig~\ref{fig:lumireact}, we show the $\stheta$ sensitivity from
reactor neutrino experiments as a function of the integrated
luminosity measured in t (fiducial far detector mass) $\times$ GW
(thermal reactor power) $\times$ yr (time of data
taking).\footnote{Note that we assume 100\% detection efficiency in the
far detector. For smaller efficiencies, one needs to re-scale the
luminosity.} We consider two options of the far detector baseline:
$L_\mathrm{FD} = 1.05\,\mathrm{km}$, corresponding to the baseline of
the CHOOZ site, and $L_\mathrm{FD} = 1.7\,\mathrm{km}$, which is
optimized for values of $\Delta m^2_{31} \sim (2-4)\cdot
10^{-3}\,\mathrm{eV}^2$~\cite{Huber:2003pm}. A crucial parameter for
the $\stheta$ sensitivity is the uncertainty of the relative
normalization between the near and far detectors. We show the
sensitivity for two representative values for this relative
normalization uncertainty: First, $\sigma_\mathrm{rel} = 0.6\%$ is a
realistic value for two identical
detectors~\cite{doubleChooz}. Second, in order to illustrate the
improvement of the performance of an reactor experiment with a reduced
normalization error, we consider the very optimistic assumption of
$\sigma_\mathrm{rel} = 0.2\%$. Such a small value might be achievable
with movable detectors, as discussed for some proposals in the
US~\cite{shaevitztalk}. The shaded regions in \Fig~\ref{fig:lumireact}
correspond to the range of possible sensitivity limits for different
assumptions of systematical errors and backgrounds. For the optimal
case (lower curves), we only include the absolute flux and relative
detector uncertainties $\sigma_\mathrm{abs}$ and
$\sigma_\mathrm{rel}$. For the worst limits (upper curves) we include
in addition an error on the spectral shape $\sigma_\mathrm{shape} =
2\%$, the energy scale uncertainty $\sigma_\mathrm{cal} = 0.5\%$, and
various backgrounds as discussed in \App~\ref{app:reactor}.

The first observation from \Fig~\ref{fig:lumireact} is that the shaded
regions in the left-hand panel are significantly wider than in the
right-hand panel, which demonstrates that a reactor experiment at $1.05
\,\mathrm{km}$ is more sensitive to systematical errors. This reflects
the fact that the baseline of $1.7\,\mathrm{km}$ is better optimized for
the used value of $|\ldm| = 2\cdot 10^{-3} \, \mathrm{eV}^2$, such that the
oscillation minimum is well contained in the center of the observed
energy range. In contrast, for the baseline of $1.05 \,\mathrm{km}$,
the signal is shifted to the low energy edge of the spectrum.
This implies that the interplay of background uncertainties, energy
calibration, and shape error has a larger impact on the
final sensitivity limit. 

However, from the left-hand panel, one finds that for experiments of
the size of \DChooz , the impact of systematics is rather modest; the
$\stheta$ sensitivity of 0.024 for normalization errors only
deteriorates to 0.032 if all systematics errors and backgrounds are included.
We conclude that the proposed \DChooz\ project is rather insensitive to
 systematical effects and will be able to provide a robust
limit $\stheta \lesssim 0.032$, although the far detector baseline is
not optimized.
In contrast, if one aims at higher luminosities, the systematics will
have to be well under control at a non-optimal baseline such as at the
CHOOZ site. In that case, it is saver to use a longer baseline. We
note that the main limiting factor for large luminosities in the
right-hand panel of \Fig~\ref{fig:lumireact} is the error on a
bin-to-bin uncorrelated background. Furthermore, comparing the light
and dark shaded regions in that plot, it is obvious that a smaller
relative normalization error will significantly improve the
performance of a large experiment at $1.7 \,\mathrm{km}$, and will
further reduce the impact of systematics and backgrounds. With the
ambitious value of $\sigma_\mathrm{rel} = 0.2\%$, sensitivity limits
of $\stheta \lesssim 7\cdot 10^{-3}$ could be obtained with a
\ReactorII -type experiment.

Eventually, we have demonstrated that the \DChooz\ experiment could
give a robust $\stheta$ sensitivity limit. In fact, one can read off
\figu{lumireact} (\CHOOZII -dot) that our assumptions about \DChooz\
are rather conservative. Since a Letter of Intent for this experiment
is in preparation, we use it in the next subsection for a direct
quantitative comparison to the superbeams. However, as one can also
learn from \figu{lumireact}, luminosity and different systematics
sources are important issues for a reactor experiment. Therefore, one
should keep in mind that much better $\stheta$ sensitivity limits
could be obtained from reactor experiments, such as the \ReactorII\
setup. However, the exact final sensitivity limits will in these cases
depend on many sources, which means that they are hardly predictable
right now.

\subsection{The $\boldsymbol{\stheta}$ bound from different 
experiments in ten years from now}
\label{sec:th13_compare}

Let us now assume that the conventional beam experiments \minos,
\icarus, and \opera\ have been running five years each, and that the
\DChooz\ experiment has accumulated three years of data. In addition,
we assume that the superbeam experiments \JHFSK\ and \NuMI\ have
reached the integrated luminosities as given in \Tab~\ref{tab:reps}.
(For earlier, more extensive discussions of the potential of superbeam
experiments, we refer to \Ref~\cite{Huber:2002rs}.)

\begin{figure}[t!]
\centering \includegraphics[width=10cm]{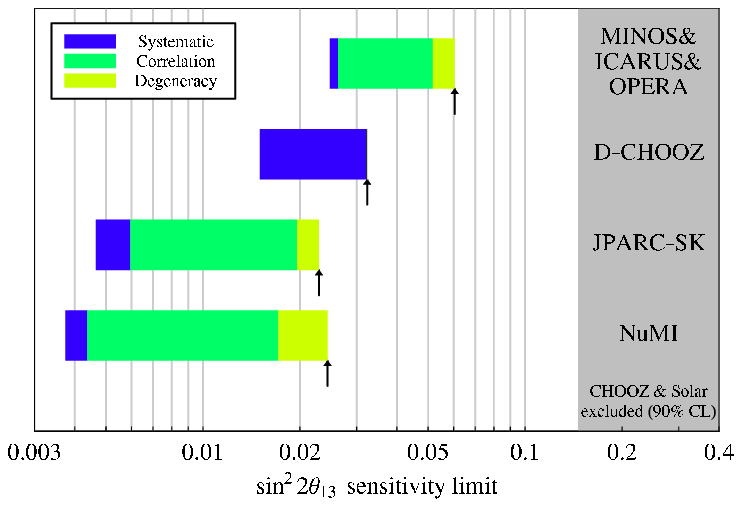}
   \mycaption{\label{fig:externcomp} The $\stheta$ sensitivity limit
   at the 90\% CL for \minos, \icarus, and \opera\ combined, \DChooz,
   \JHFSK, and \NuMI. The left edges of the bars are obtained for the
   statistics limits only, whereas the right edges are obtained after
   successively switching on systematics, correlations, and
   degeneracies, \ie, they correspond to the final $\stheta$ sensitivity
   limits. The gray-shaded region corresponds to the current
   $\sin^22\theta_{13}$ bound at 90\% CL. For the true values of the
   oscillation parameters, we use the current best-fit values in
   \eq~(\ref{eq:bfp}) and a normal mass hierarchy.}
\end{figure}

In \figu{externcomp}, we show the $\stheta$ sensitivity for the considered
experiments.  The final sensitivity limit is obtained after
successively switching on systematics, correlations, and degeneracies
as the rightmost edge of the bars.\footnote{Note that earlier similar
figures, such as in \Refs~\cite{Huber:2002rs,Huber:2003pm}, are
computed with different parameter values, which leads to changes of
the final sensitivity limits. The largest of these changes come from the
adjusted atmospheric best-fit values and \NuMI\ parameters.}
\figu{externcomp} demonstrates that the beam experiments are dominated
by correlations and degeneracies, whereas the reactor experiments are
dominated by systematics. It can be clearly seen that the
$(\stheta)_\mathrm{eff}$ sensitivity limit (between systematics and
correlation bar), or the precision of a combination of parameters
leading to a positive signal, is much better for the superbeams than
for the reactor experiments. Therefore, though the reactor experiments
have a good potential to extract $\stheta$ directly, the superbeams
results will in addition contain a lot of indirect information about
$\deltacp$ and the mass hierarchy, which might be resolved by the
combination with complementary information. We call this gain in the
physics potential which goes beyond the simple addition of statistics
for the combination of experiments ``synergy''. In
\Sec~\ref{sec:th13delta}, we will discuss this further for the case if
$\stheta$ turns out to be large.

Another conclusion from \figu{externcomp} is that it is very important
to compare the $\stheta$ sensitivities of different experiments which
are obtained with equal methods. In particular one clearly has to
distinguish between the $(\stheta)_\mathrm{eff}$ sensitivity limit
(between systematics and correlation bar) and the final $\stheta$
sensitivity limit, including correlations and degeneracies. For
example, by accident the $(\stheta)_\mathrm{eff}$ sensitivity limit
from the combined \minos , \icarus , and \opera\ experiments is very
close to the final sensitivity limit of \JHFSK\ or \NuMI. Thus, one
may end up with two similar numbers, which however, refer to different
quantities and are not comparable. 

\begin{figure}[t!]
   \centering \includegraphics[width=15.5cm]{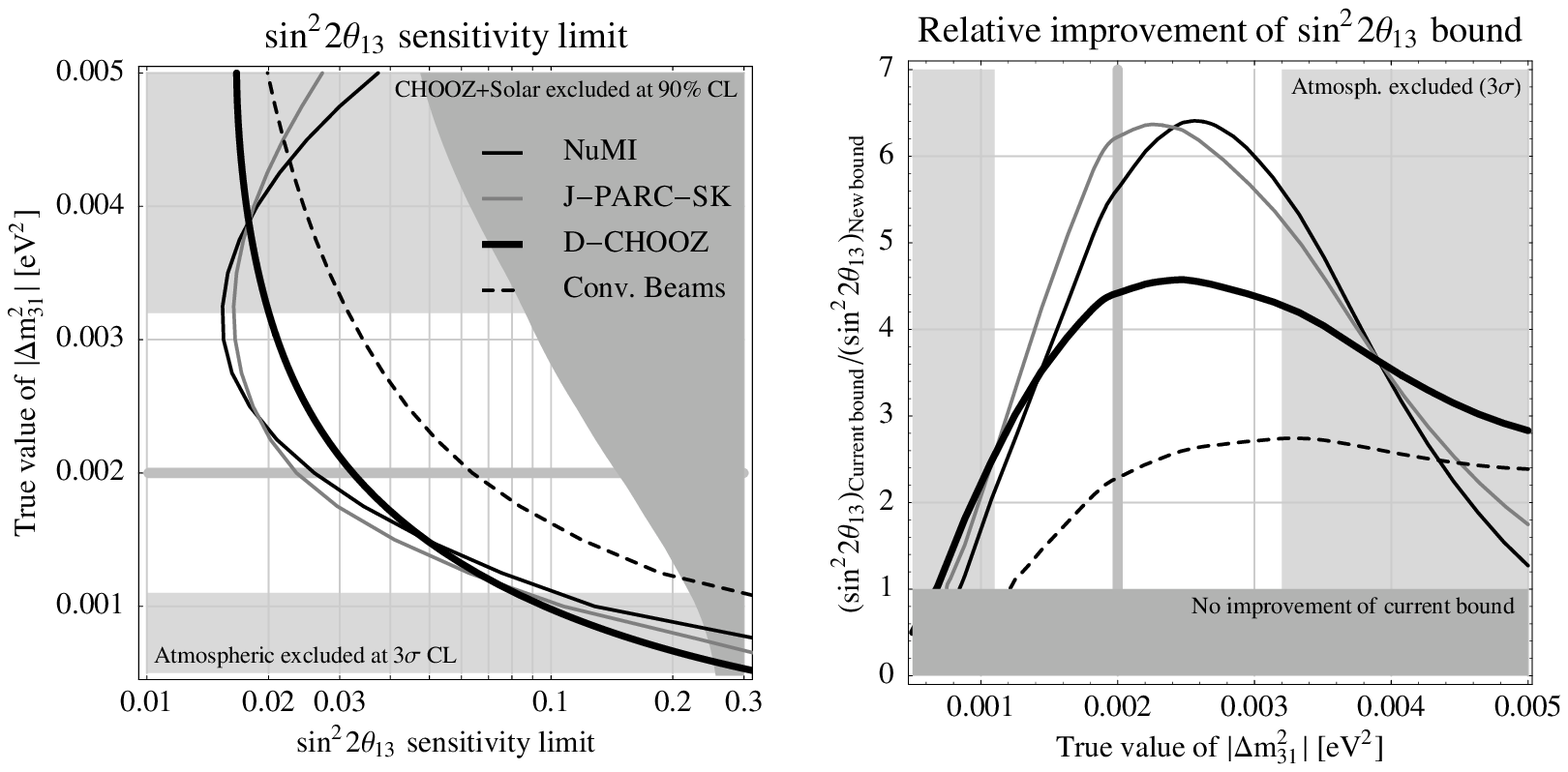}
   \mycaption{\label{fig:th13dmdep} Left panel: The $\stheta$
   sensitivity limits at 90\% CL from the experiments \NuMI, \JHFSK,
   \DChooz, and the combined conventional beams (\minos, \icarus,
   \opera) as function of the true value of $| \ldm |$.  The dark-gray
   shaded region refers to the current $\stheta$ bound from CHOOZ and
   the solar experiments (90\% CL)~\protect\cite{Maltoni:2003da}. Right panel:
   The relative improvement of the $\stheta$ sensitivity limit with
   respect to the current bound from CHOOZ and solar experiments,
   where the dark-gray region corresponds to no improvement. The
   light-gray shaded regions in both panels refer to the atmospheric
   excluded regions ($3 \sigma$), and the lines in the middle mark
   the current atmospheric best-fit value.} 
\end{figure}

A very important parameter for future $\stheta$ measurements is the
true value of $\ldm$, which currently is constrained to the interval
$0.0011 \,\mathrm{eV}^2 \lesssim | \ldm | \lesssim 0.0032
\,\mathrm{eV}^2$ at $3 \sigma$~\cite{SKupdate}. From \figu{th13dmdep},
one can can easily see that the true value of $\ldm$ strongly affects
the $\stheta$ sensitivity limit. The left-hand plot in this figure
demonstrates that for all experiments the $\stheta$ sensitivity
becomes worse for small values of $|\ldm|$ within the currently
allowed range. However, since also the current $\stheta$ bound
(dark-gray shaded region) is worse for small values of $|\ldm|$ than
for large values, the relative improvement of the current $\stheta$
bound might be a more appropriate description of the experiment
performance. This relative improvement as function of the true value
of $|\ldm|$ is shown in the right-hand plot of \figu{th13dmdep}, where
a factor of unity corresponds to no improvement. From this plot, one
can read off an improvement by a factor of two for the conventional
beams, a factor of four for \DChooz , and a factor of six for the
superbeams at the atmospheric best-fit value (vertical line).
Nevertheless, the conventional beams might not improve the current
bound at all for small values of $| \ldm |$ within the atmospheric
allowed range, whereas any of the other experiments would improve the
current bound at least by a factor of two. Thus, though the $\stheta$
sensitivity limit could be as be as large as $\stheta \lesssim 0.1$ for
small values of $| \ldm |$ for the superbeams or \DChooz, those
experiments would still improve the current bound by a factor of two. 

%%%%%%%%%%%%%%%%%%%%%%%%%%%%%%%%%%%%%%%%%%%%%%%%%%%%%%%%%%%%%%%%%%%%%%%%%%%%%%
%%%%%%%                  RESULTS IV                             %%%%%%%%%%%%%%
%%%%%%%%%%%%%%%%%%%%%%%%%%%%%%%%%%%%%%%%%%%%%%%%%%%%%%%%%%%%%%%%%%%%%%%%%%%%%%

\section{Opportunities if $\boldsymbol{\stheta}$ is just around the corner}
\label{sec:th13delta}

In \Sec~\ref{sec:theta13_all}, we have discussed how much the
$\stheta$ bound could be improved if the true value of $\stheta$ were
zero. There are, however, very good theoretical reasons to expect
$\stheta$ to be finite, such that the experiments under consideration
could establish $\stheta>0$. In this case, one could aim for the
$\stheta$ precision, CP violation, CP precision measurements, and the
mass hierarchy determination. Though it has been shown that CP and
mass hierarchy measurements are very difficult for the
first-generation superbeams and new reactor
experiments~\cite{Minakata:2002jv,Huber:2002mx,Huber:2002rs,Huber:2003pm,Winter:2003ye},
we will demonstrate in this section that we could still learn
something about these parameters if $\stheta$ turns out to be
large. In particular, we discuss the combination of the discussed
experiments for the case $\stheta=0.1$. This would imply that a
positive $\stheta$ signal could already be seen with the next
generation of experiments.
As discussed in \Sec~\ref{sec:expclasses}, we assume here that a
large reactor experiment \ReactorII\ will be available at the end of the
period under consideration to resolve the
correlation between $\stheta$ and $\deltacp$. We note again that
similar results can be obtained by the superbeams in the antineutrino
mode using higher target powers or detector upgrades.\footnote{In
fact, one could already obtain some CP-conjugate information by
running \NuMI\ at $L=712 \, \mathrm{km}$ with antineutrinos
only~\cite{Winter:2003ye}. However, we do not consider an option with
a very extensive {\em a priori} \NuMI\ antineutrino running in this
study, since the risk of this configuration is too high as long as
$\stheta>0$ is not established.}

\begin{figure}[t!]
\centering \includegraphics[width=16cm]{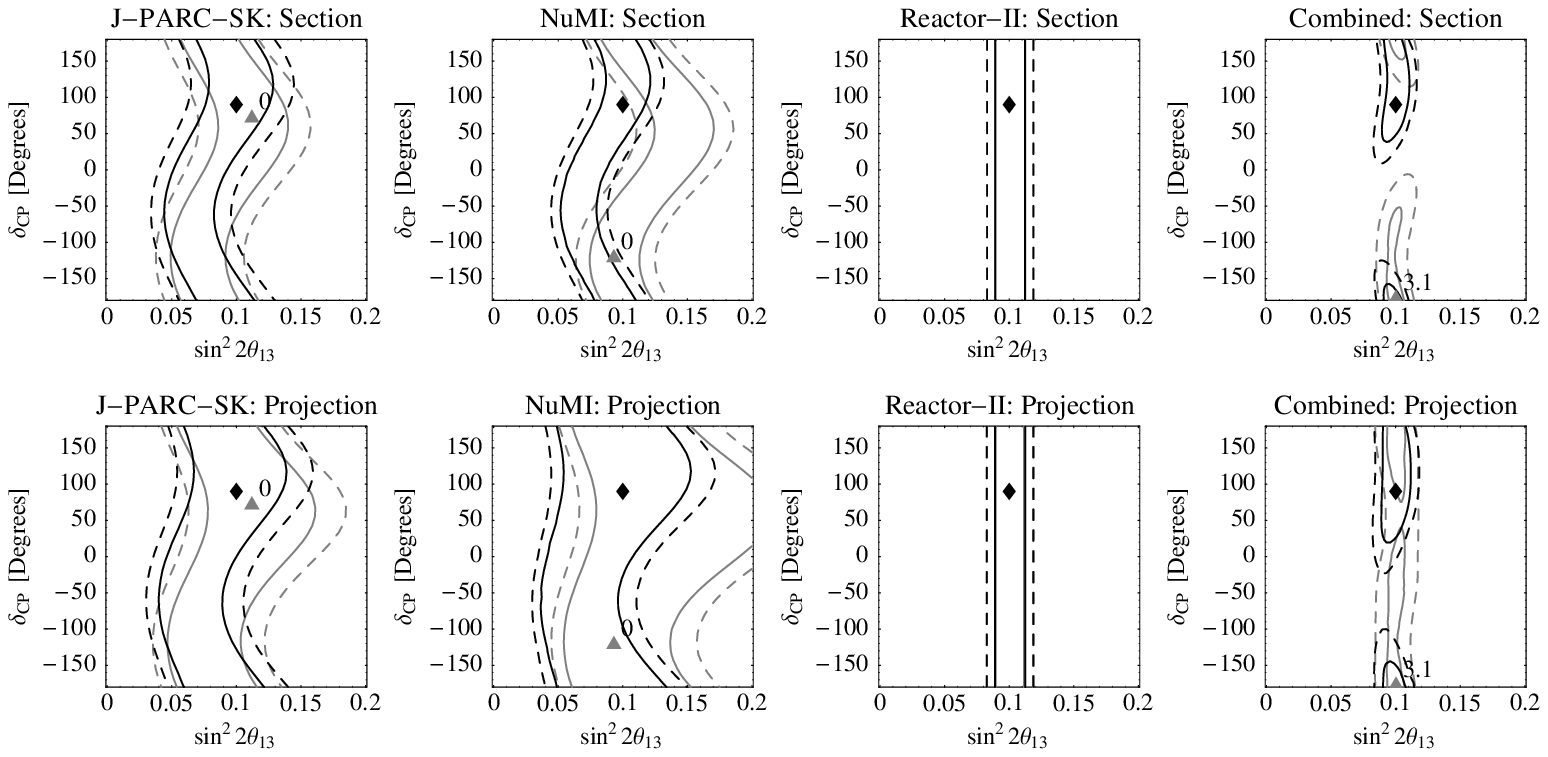}
   \mycaption{\label{fig:deltatheta1} The $90 \%$ CL (solid curves)
   and $3 \sigma$ (dashed curves) allowed regions (2 d.o.f.) in the
   $\stheta$-$\deltacp$-plane for the true values $\stheta=0.1$ and
   $\deltacp=90^\circ$ for \JHFSK, \NuMI, \ReactorII. The right-most
   plots are calculated for the shown experiments in combination with the
   conventional beams. For the true values of the un-displayed
   oscillation parameters, we choose the current best-fit values and a
   normal mass hierarchy. The black curves refer to the allowed
   regions for the normal mass hierarchy, whereas the gray curves
   refer to the $\mathrm{sgn}(\ldm)$-degenerate solution (inverted
   hierarchy), where the projections of the minima onto the
   $\stheta$-$\deltacp$-plane are shown as diamonds (normal hierarchy)
   and triangles (inverted hierarchy). For the latter, the
   $\Delta\chi^2$-value with respect to the best-fit point is also
   given. The upper row shows the fit manifold section (with the
   un-displayed oscillation parameters fixed at their true values), and
   the lower row shows the projection onto the
   $\stheta$-$\deltacp$-plane as the final result.}
\end{figure}

\begin{figure}[t!]
\centering
\includegraphics[width=16cm]{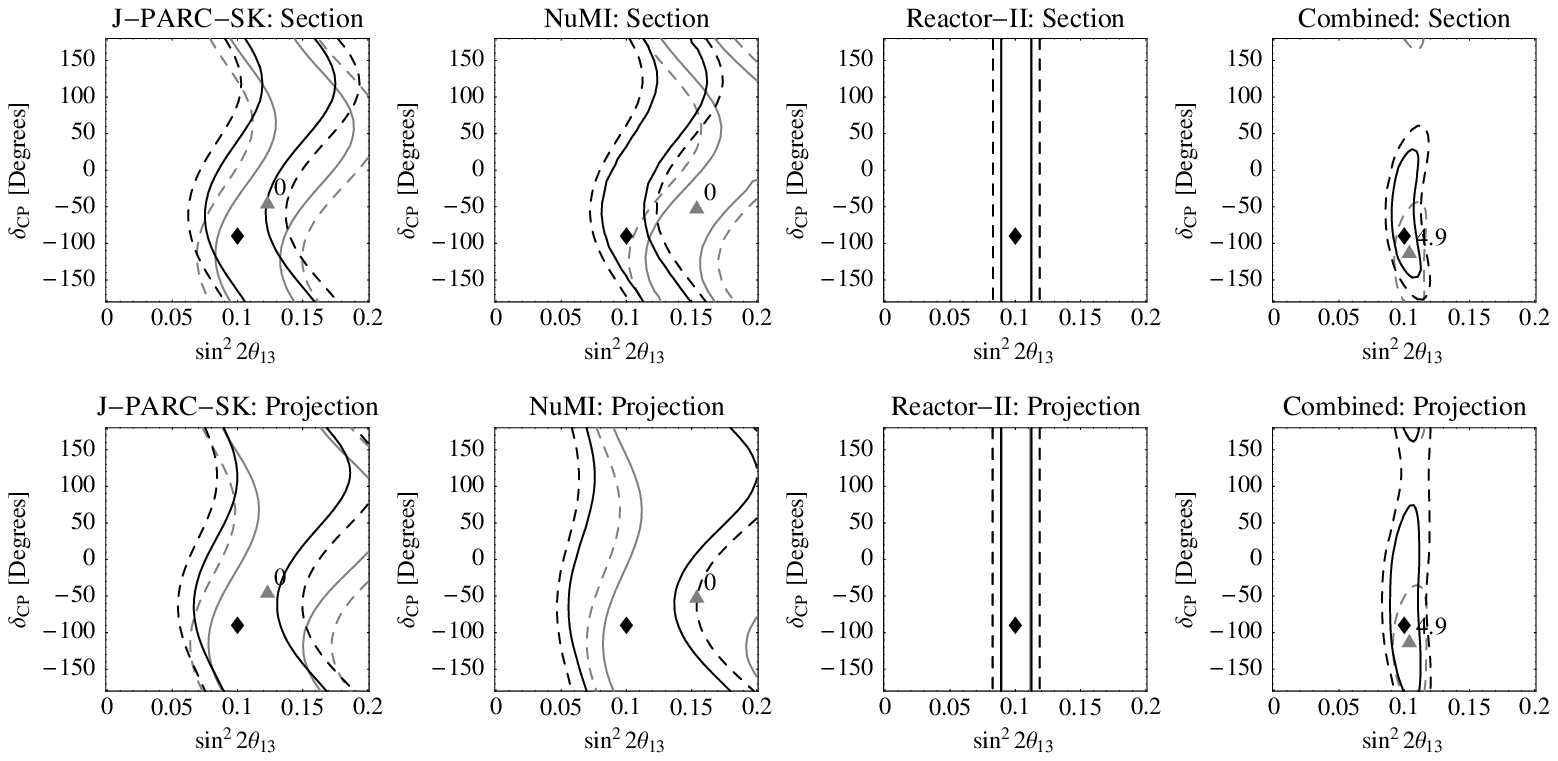}
\mycaption{\label{fig:deltatheta2} The same as \figu{deltatheta1} but
  for the true value $\deltacp=-90^\circ$.} 
\end{figure}

The superbeam appearance channels will lead to allowed regions in the $\stheta$-$\deltacp$-plane,
similar to the allowed regions for solar and atmospheric oscillation
parameters from current data. We show the results of \JHFSK, \NuMI,
and \ReactorII\ for the true values $\stheta=0.1$ and
$\deltacp=90^\circ$ in \fig~\ref{fig:deltatheta1}, and
$\deltacp=-90^\circ$ in \fig~\ref{fig:deltatheta2}. For the right-most
plots in these figures, we combine all experiments including the
conventional beams \minos, \icarus, and \opera, although they do not
contribute significantly to the final result. Since we assume a normal
mass hierarchy to generate the data, the best-fit is obtained by
fitting with the normal hierarchy; the corresponding regions are shown
by the black curves. The $\mathrm{sgn}(\ldm)$-degenerate regions are
obtained by fitting the data assuming an inverted hierarchy (gray curves). 
Thus, the best-fit and degenerate manifolds,
which are disconnected in the six-dimensional parameter space, are
shown in the same plots. Similar to \Fig~\ref{fig:dm31th23} we
demonstrate the difference between a section of the fit manifold
(upper rows) and a projection (lower rows) in these figures.

As far as the measurement of $\stheta$ is concerned, any of the
experiments in \figs~\ref{fig:deltatheta1} and~\ref{fig:deltatheta2}
can establish $\stheta>0$ for $\stheta=0.1$ at $3\sigma$.  The
$\stheta$-precision can be read off from the figures as projection of the
bands onto the $\stheta$-axis.\footnote{Note that these figures are
computed for two degrees of freedom, which means that the projections
with one degree of freedom are slightly smaller. In fact, the $90 \%$
CL contour for two degrees of freedom ($\Delta \chi^2=4.61$) is close
to the $2 \sigma$ contour for one degree of freedom ($\Delta
\chi^2=4.00$). In particular, for the sake of comparison, we also use
2~d.o.f.\ for \ReactorII, although it does not depend on $\deltacp$.}
The band structures of \JHFSK\ and \NuMI\ come from the CP phase
dependency in \equ{beam}.  Because of the larger matter effects, the
degenerate solution for \NuMI\ is rather different from the best-fit
solution, whereas it is very similar to the best-fit solution for
\JHFSK. For \ReactorII, the $\stheta$-precision can be read off
directly, since a reactor experiment is not affected by $\deltacp$
(see \equ{reactor}), and the mass hierarchy has essentially no
effect. Note that the treatment of the $\mathrm{sgn}(\ldm)$-degeneracy
in such a situation as shown in \figs~\ref{fig:deltatheta1}
 and~\ref{fig:deltatheta2} is a matter of definition: One could either 
return two different
intervals for normal and inverted mass hierarchies, or one could
return the union of the two fit intervals as more condensed
information.

The figures show that for $\deltacp$, none of the individual
experiments can give any information, since no substantial fraction of
antineutrino running is involved. However, there is some information
on $\deltacp$ for the combination of all experiments, since the
complementary information from the reactor experiment helps to resolve
the superbeam bands. Note that the overall performance for the
considered experiments (including the $\mathrm{sgn}(\ldm)$-degeneracy)
is usually better close to the true value $\deltacp=-90^\circ$ than
close to the true value $\deltacp=90^\circ$, since the degeneracy
includes for $\deltacp=90^\circ$ very different values of $\deltacp$
far away from the best-fit manifold. This can, for example, be
understood in terms of bi-rate graphs (\cf , \Refs~\cite{Minakata:2001qm,Winter:2003ye}). From a separate analysis of the CP
precision, we find that one could exclude as much as up to 40\%
of all values of $\deltacp$ at the $90\%$ confidence level (1 d.o.f., close to
$\deltacp=-90^\circ$). However, if $\deltacp$ turns out to be close to
$0$ or $\pi$, we find that one could not obtain any information on
$\deltacp$. Furthermore, one can directly read off from
\figs~\ref{fig:deltatheta1} and~\ref{fig:deltatheta2} that CP
violation measurements will not be possible with the considered
experiments at the $90\%$ confidence level (2~d.o.f. in these
figures), since for the true values $\deltacp=\pm 90^\circ$
corresponding to maximal CP violation, the projected allowed regions
(even the best-fit solutions) include at least one of the CP
conserving cases $\deltacp \in \{0, 180^\circ\}$. One can show that
even for one degree of freedom, there is no CP violation sensitivity
with the discussed experiments at the $90\%$ confidence level.

Another important issue for the next generation long-baseline
experiments is the mass hierarchy determination. In
\figs~\ref{fig:deltatheta1} and~\ref{fig:deltatheta2} we give the
$\Delta\chi^2$-values for the minimum in the fit manifold
corresponding to the $\mathrm{sgn}(\ldm)$-degenerate solution (\ie,
for the inverted mass hierarchy) with respect to the best-fit minimum
for the normal hierarchy ($\Delta \chi^2=0$), which is the relevant
number for the sensitivity to a normal mass hierarchy. Obviously, none
of the individual experiments has a mass hierarchy sensitivity, but
their combination has some. The mass hierarchy sensitivity becomes only
possible because of the long \NuMI\ baseline $L=812 \,
\mathrm{km}$~\cite{Barger:2002xk,Huber:2002rs,Minakata:2003ca}, since
matter effects differ for the normal and inverted mass hierarchies.
Eventually, a \NuMI\ baseline even longer than $L=812 \, \mathrm{km}$
could further improve the mass hierarchy
sensitivity~\cite{Huber:2002rs,Winter:2003ye}.
We note that the ability to identify the mass hierarchy strongly
depends on the (unknown) true value of $\deltacp$. The mass hierarchy
determination at the combined superbeams is close to the optimum for
$\deltacp=-90^\circ$, and close to the minimum for
$\deltacp=90^\circ$~\cite{Minakata:2001qm,Winter:2003ye}.  In fact, one could have
a better sensitivity to the normal mass hierarchy for $\deltacp=-90^\circ$
($\Delta \chi^2 = 4.9$, see \figu{deltatheta2}) than for $\deltacp=90^\circ$ ($\Delta \chi^2 = 3.1$, see \figu{deltatheta1}) at the $90\%$ confidence level ($\Delta \chi^2 = 2.71$ for 1~d.o.f.).

%%%%%%%%%%%%%%%%%%%%%%%%%%%%%%%%%%%%%%%%%%%%%%%%%%%%%%%%%%%%%%%%%%%%%%%%%%%%%%
%%%%%                  CONCLUSIONS                              %%%%%%%%%%%%%%
%%%%%%%%%%%%%%%%%%%%%%%%%%%%%%%%%%%%%%%%%%%%%%%%%%%%%%%%%%%%%%%%%%%%%%%%%%%%%%

\section{Summary and conclusions}
\label{sec:conclusions}

This study has focused on the future neutrino oscillation
long-baseline experiments on a timescale of about ten years.  The
primary objective has been the search for $\stheta$, but we have also
analyzed the ``atmospheric'' parameters $\theta_{23}$ and $\ldm$.  The
main selection criterion for the different experiments has been the
availability of specific studies, such as LOIs or proposals, or that they
are even being under construction. We
assume that an experiment (including data taking and analysis) will only be
feasible within the coming ten years, if it is already now actively
being planned. The next long-baseline experiments will be the
conventional beam experiments \minos , \icarus , and \opera\, which
are currently under construction. In addition, the \JHFSK\ and \NuMI\
superbeam experiments are under active consideration with existing
proposals and will most likely provide results within the next ten
years. Furthermore, new reactor neutrino experiments are
actively being discussed. In this study, we have considered the \DChooz\
project, which will probably deliver results in
the anticipated timescale, since infrastructure (such as the detector
cavity) of the original CHOOZ experiment can be re-used.

First, we have investigated the possible improvement of our knowledge
on the leading atmospheric oscillation parameters. We have found that
the conventional beams and superbeams will reduce the error on $\ldm$ 
by roughly an order of magnitude within the next ten years. The precision of $\theta_{23}$ is dominated by \JHFSK\ and will improve only by a factor of two (\cf, \Tab~\ref{tab:summary}).

\begin{table}[t!]
\begin{center}
{\small
\begin{tabular}{|lrrrrrrr|}
\hline
& Current & Beams & \CHOOZII\ & \JHFSK\ & \NuMI\ & \ReactorII\ & Comb. \\  
\hline
\multicolumn{8}{|l|}{$\stheta$ sensitivity limit (90\% CL)}
 \\
\hline
\\[-0.4cm]
$\stheta$ &  0.14 & 0.061 & 0.032 & {\bf 0.023} & 0.024 & (0.009) & (0.009) \\[0.1cm]
$(\stheta)_{\mathrm{eff}}$ & 0.14 & 0.026 & 0.032 & 0.006 & {\bf 0.004} & (0.009) & (0.003) \\[0.05cm]
\hline
\multicolumn{8}{|l|}{Allowed ranges for leading atmospheric parameters ($3 \sigma$)} \\
\hline
\\[-0.4cm]
$\frac{|\ldm|}{10^{-3} \, \mathrm{eV}^2}$ 
& 2$^{+1.2}_{-0.9}$ & 2$^{+0.34}_{-0.18}$ & $-$ &  2$^{\boldsymbol{+0.15}}_{-0.09}$ & 2$^{+0.43}_{\boldsymbol{-0.07}}$ & $-$ & 2$^{+0.12}_{-0.06}$ \\[0.1cm]
$\theta_{23}$ & $(\frac{\pi}{4})^{+0.20}_{-0.20}$  & $(\frac{\pi}{4})^{+0.22}_{-0.19}$ & $-$ & $(\frac{\pi}{4})\boldsymbol{^{+0.13}_{-0.10}}$ & $(\frac{\pi}{4})^{+0.24}_{-0.21}$ & $-$ & $(\frac{\pi}{4})^{+0.13}_{-0.10}$\\[0.1cm]
\hline
\multicolumn{8}{|l|}{Measurements for large $\stheta=0.1$ (90\% CL)} \\
\hline
\\[-0.4cm]
$\stheta$ & $-$ & 0.1$^{+0.104}_{-0.052}$ & 0.1$^{+0.034}_{-0.033}$ & 0.1$^{+0.067}_{-0.034}$ & 0.1$^{+0.083}_{-0.043}$ & 0.1$^{\boldsymbol{+0.010}}_{\boldsymbol{-0.008}}$ & 0.1$^{+0.010}_{-0.008}$ \\[0.1cm] 
$\deltacp$ & 
\multicolumn{7}{l|}{Combination can exclude up to 40\% of all values} \\[0.05cm]
CP violation & \multicolumn{7}{l|}{No sensitivity to CP violation of any tested experiment or combination} \\[0.05cm]
$\mathrm{sgn}(\ldm)$ & \multicolumn{7}{l|}{Combination has sensitivity to normal mass hierarchy} \\[0.05cm]
\hline
\end{tabular}
} % small
\end{center}
\mycaption{\label{tab:summary} Summary table of this study.  The
   numbers which are printed boldface represent the best individual
   results within each row. For the true values of the oscillation
   parameters, we use the current best-fit values from
   \eq~(\ref{eq:bfp}) and a normal mass hierarchy. The precisions for
   $\stheta$ do not include the $\mathrm{sgn}(\ldm)$-degeneracy and are
   computed for the true value $\deltacp=0$. If one
   does not use \ReactorII\ for the combination of all experiments,
   but \DChooz\ instead, one obtains the following values: 0.016 for
   the $\stheta$ limit, 0.003 for the $(\stheta)_{\mathrm{eff}}$
   limit, and 0.1$^{+0.025}_{-0.021}$ for the $\stheta$ precision.}
\end{table}

As the next important issue, we have investigated the potential of the
conventional beams, \ie , \minos , \icarus , and \opera , to improve
the current $\stheta$ bound from CHOOZ and the solar experiments in a
complete analysis taking into account correlations and degeneracies.
Since the final luminosities of these experiments are not yet
determined, we have discussed the results as function of the total
number of protons on target.
We have found that \minos\ could improve the current bound after a
running time of about two years, and \icarus\ and \opera\ combined
after about one and a half years. In addition, we have discussed the
maximal potential of all three conventional beams combined with a running
time of five years each, leading to a final sensitivity
limit of $\stheta \le 0.061$ (all sensitivity limits at 90\% CL). This final
sensitivity limit includes correlations and degeneracies, which means
that it reflects the experiment's ability to extract the parameter
$\stheta$ from the appearance information. Since correlations and
degeneracies could be reduced by later experiments, another
interesting measure is the systematics-only $\stheta$ limit for fixed
oscillation parameters, \ie, the sensitivity limit to a specific
combination of parameters, which we have called
$(\stheta)_{\mathrm{eff}}$.
We have found a $(\stheta)_{\mathrm{eff}}$ sensitivity limit for the
conventional beams of 0.026, illustrating that correlations and
degeneracies have a rather large impact on the $\stheta$ limit from
conventional beams.
Note that it is important to compare different experiments with equal
methods, which means that only $\stheta$ or $(\stheta)_{\mathrm{eff}}$
limits should be compared with each other. In addition, it is
interesting to observe that the final $\stheta$ sensitivity limit
increases with increasing $\sdm$ within the solar allowed region,
whereas the $(\stheta)_{\mathrm{eff}}$ sensitivity limit
decreases. This can be understood by the amplitude of the
$\deltacp$-terms which is proportional to $\sdm$.  

Furthermore, we have investigated the $\stheta$-limit obtainable by
nuclear reactor experiments. A thorough analysis of the \DChooz\ configuration
including systematics and backgrounds, demonstrates that a robust
limit of $\stheta \le 0.032$ can be obtained in spite of the non-optimal
baseline of $1.05\,\mathrm{km}$. If one aims, however, to 
significantly higher luminosities than the $60\,000$ events anticipated by \DChooz, the systematics has to be well under control. In this case,
a more optimized baseline of $1.7\,\mathrm{km}$ helps to reduce the
impact of systematics and backgrounds, and limits of the order of $\stheta \le 0.014$ could be achievable.

If in ten years from now no finite value is established, $\stheta$
bounds from the conventional beams (\minos, \icarus, \opera), from
reactor experiments, such as \DChooz, and from the superbeams \JHFSK\ and
\NuMI\ will be available. We have demonstrated that the conventional
beams could improve the current $\stheta$ bound by about a factor of
two, the \DChooz\ experiment by about a factor of four, and the
superbeams by about a factor of six. We have also shown that these
results apply to a large range within the allowed interval for $\ldm$,
since not only the experiment's potential decreases for small values
of $\ldm$, but also the current $\stheta$ bound. For $\ldm = 2\cdot
10^{-3} \,\mathrm{eV}^2$ we have found a final $\stheta$ sensitivity
limit of $\stheta \le 0.02$ for the superbeams.  Note that, though the \DChooz\
setup is not as good as the superbeams, its results are not affected
by the true value of $\sdm$ within the solar-allowed
range~\cite{Huber:2003pm}, which means that a reactor experiment is
more robust with respect to the true parameter values.  Moreover,
because correlations and degeneracies do not effect the $\stheta$ limit
from reactor experiments, the $\stheta$ and $(\stheta)_{\mathrm{eff}}$
limits are almost identical for \DChooz. In contrast, for the superbeams the
$(\stheta)_{\mathrm{eff}}$ limit is nearly one order of magnitude
smaller than the $\stheta$ limit (\cf, \Tab~\ref{tab:summary}),
demonstrating that correlations and degeneracies are crucial for them.

In order to illustrate where we could stand in ten years from now if
$\stheta$ were close to the current bound, we have also performed an analysis by assuming $\stheta = 0.1$. In this case,
 all the considered experiments will establish the
finite value of $\stheta$ and measure it with a certain precision (\cf,
\Tab~\ref{tab:summary}). In this situation, which is theoretically well
motivated (\cf, \Tab~1 of \Ref~\cite{whitepaper}), one could even aim
to learn something about $\deltacp$ and the neutrino mass hierarchy
with the next generation of experiments.
Since the results of superbeam experiments will lead to strong
correlations between $\deltacp$ and $\stheta$, it is well known that
complementary information is needed to disentangle these two
parameters. One can either use extensive antineutrino running at the
superbeams (which, however might not be possible at the time scale of
ten years, because of the lower antineutrino cross sections), or a
large reactor experiment to measure $\stheta$. In this study, we have
demonstrated the potential for $\deltacp$ by assuming such a large
reactor experiment at an ideal baseline of $L=1.7 \, \mathrm{km}$,
which we call \ReactorII. Though such an experiment might not 
exactly fit into the discussed timescale, it might be realized soon
thereafter. Possible sites for such an experiment are under
investigation~\cite{whitepaper} (some proposals, which are, for
example, discussed in the US, are similar to our \ReactorII\ setup).
Indeed, we find that in this optimal situation ($\stheta=0.1$), up to
40\% of all possible values for $\deltacp$ could be
excluded (90 \% CL). This result, however, depends strongly on the
true value of $\deltacp$, and applies to maximal CP violation. For the case
of CP conservation (true parameter value), however, nothing at all could be learned about $\deltacp$. In either case, a sensitivity to CP violation would not be
achievable with the discussed experiments because of too low
statistics. For the mass hierarchy determination, we have found that
one would be sensitive to a normal mass hierarchy at the 90\% confidence
level, where the sensitivity to the
inverted mass hierarchy would be somewhat
worse~\cite{Winter:2003ye}. Note that the sensitivity to the mass
hierarchy is mainly determined by matter effects in \NuMI, and could
even be better for \NuMI-baselines larger than $812\,\mathrm{km}$.

To summarize, from the current perspective neutrino oscillations will
remain a very exciting field of research, and the experiments
considered within the next ten years will significantly improve our
knowledge. Eventually, these experiments could indeed restrict
$\deltacp$ and determine the neutrino mass hierarchy within the coming ten to
fifteen years if $\stheta$ turns out to be sizeable. The remaining
 ambiguities could be resolved by the subsequent generation of experiments, 
such as superbeam upgrades, beta beams, or neutrino
factories~\cite{Barger:2002xk,Wang:2001ys}. 
 
%%%%%%%%%%%%%%%%%%%%%%%%%%%%%%%%%%%%%%%%%%%%%%%%%%%%%%%%%%%%%%%%%%%%%%%%%
%               ACKNOWLEDGMENTS                             %%%%%%%%%%%%
%%%%%%%%%%%%%%%%%%%%%%%%%%%%%%%%%%%%%%%%%%%%%%%%%%%%%%%%%%%%%%%%%%%%%%%%%

\subsubsection*{Acknowledgments}

We want to thank M.~Goodman and K.~Lang for providing input data and
useful information about the MINOS experiment, especially the PH2low
beam flux data.  Furthermore, we thank T.~Lasserre, H.~deKerret and
L.~Oberauer for very useful discussions on the Double-Chooz project.
This study has been supported by the ``Sonderforschungsbereich 375
f{\"u}r Astro-Teilchenphysik der Deutschen
For\-schungs\-ge\-mein\-schaft''.

\newpage

\begin{appendix}

\section{Simulation details of the conventional beam experiments}
\label{sec:simbeams}

In this appendix, we describe our simulations of the conventional beam
experiments \minos, \icarus, and \opera\ in greater detail. In
\App~\ref{app:parameters} we give the numbers and references for the
experimental parameters used, and in \App~\ref{app:reproduction} we
demonstrate that our calculations reproduce the results of the
simulations of the experimental collaborations to good accuracy.

\subsection{Description of the experiments and experimental parameters}
\label{app:parameters}

The \minos\ experiment will use both a near and far detector.  The
near detector allows to measure the neutrino flux and energy spectrum.
In addition, other important characteristics, such as the initial
$\nu_e$ contamination of the un-oscillated neutrino beam can be
extracted with good precision. Besides the smaller detector mass of
$1\,\mathrm{kt}$, it is constructed as identical as possible to the
far detector in order to suppress systematical uncertainties.  The far
detector is placed $713\,\mathrm{m}$ deep in a newly built cavern in
the Soudan mine in order to suppress cosmic ray backgrounds. It is an
octagonal, magnetized iron calorimeter with a diameter of
$8\,\mathrm{m}$, assembled of steel layers alternating with
scintillator strips with an overall mass of $5.4\,\mathrm{kt}$. The
construction of the far detector was finished in spring of 2003
and it is now taking data on atmospheric neutrinos and
muons. 

The mean energy of the neutrino beam produced at Fermilab can be
varied between $3$ and $18\,\mathrm{GeV}$. The beam is planned to
start with the low energy configuration (PH2low), with the peak
neutrino energy at $\langle E_\nu \rangle \sim 3\,\mathrm{GeV}$.  In
our simulation we use the official PH2low beam
configuration~\cite{NUMI714}, which means that we do not include a
hadronic hose or different beam-plugs in the beam line setup.  These
modifications would lead to a better signal to background
ratio. However, as discussed in \Ref~\cite{NUMI714}, they affect the
sensitivity limits to $\sin^22\theta_{13}$ only marginally. The NuMI
PH2low beam flux data, as well as the detection cross sections have
been provided by \Ref~\cite{MG}. We use $30$ energy bins in the energy
range between $2\,\mathrm{GeV}$ and $6\,\mathrm{GeV}$. In addition,
the energy resolution is assumed to be $\sigma_E=0.15\cdot
E_\nu$~\cite{Ables:1995wq}. The NuMI beam will have a luminosity of
$3.7\cdot10^{20}\,\mathrm{pot}\,\mathrm{y}^{-1}$. In addition to the
$\nu_\mu\to\nu_e$ appearance channel most relevant for the $\stheta$
measurement, we include also the $\nu_\mu\rightarrow\nu_\mu$
disappearance channel with an efficiency of $0.9$, and we take into
account that a fraction of $0.05$ of the neutral current background
events will be misidentified as signal events.

For the CNGS experiments, we use the flux and cross sections from
\Ref~\cite{CNGSflux}. For both \icarus\ and \opera , we use an energy
range between $1$ and $30\,\mathrm{GeV}$, which is divided into $80$
bins. For \icarus , we assume an energy resolution of
$\sigma_E=0.1\cdot E_\nu$~\cite{Komatsu:2002sz}, and for \opera\
$\sigma_E=0.25 \cdot E_\nu$. The latter might be somewhat
overestimated~\cite{Duchesneau:2002yq}. However, our $\theta_{13}$
limit at \opera\ changes less than $5\%$ for values of the energy
resolution up to $\sigma_E=0.4 \cdot E_\nu$.  For the CNGS beam, we
assume a nominal luminosity of
$4.5\cdot10^{19}\,\mathrm{pot}\,\mathrm{y}^{-1}$.

The original purpose of the CNGS experiments is the observation of
$\nu_\mu\to \nu_\tau$ appearance.  The \opera\ detector is an emulsion
cloud chamber, and the extremely high granularity of the emulsion
allows to detect the $\nu_\tau$ events directly by the so-called
``kink'', which comes from the semi-leptonic decay of the tauons.  In
order to reach a significant detector mass, the emulsion layers are
separated by lead plates of $1\,\mathrm{mm}$ thickness. The total
fiducial mass of the detector will be $1.8\,\mathrm{kt}$. However,
during the extraction of the data, the detector mass will change as a
function of time. Therefore, we use the time averaged fiducial mass of
$1.65\,\mathrm{kt}$ for our analysis.  A main challenge in the \opera\
experiment is the automated scanning of the emulsions. The \icarus\
detector uses a different approach: it is a liquid Argon TPC, which
allows to reconstruct the three dimensional topology of an event with
a spacial resolution of roughly $1\,\mathrm{mm}$ on an event by event
basis. The $\nu_\tau$ detection is performed by a full kinematical
analysis.  The fiducial mass will be $2.35 \,\mathrm{kt}$.

For the \opera\ experiment, we include the information from the
$\nu_\mu\to\nu_\tau$ channel by assuming an efficiency of $0.11$, and
a fraction of $3\cdot10^{-5}$ of misidentified neutral current events.
For the \icarus\ experiment, we use an efficiency of $0.075$ for this
channel with a background fraction of $8.5\cdot10^{-5}$ of the neutral
current events~\cite{Aprili:2002wx}.
Although the \icarus\ and \opera\ detectors are
optimized to observe the decay properties of $\tau$-leptons, they also
have very good abilities for muon identification, which allows to 
measure also $\nu_\mu$ disappearance. We therefore include
the $\nu_\mu\rightarrow\nu_\mu$ CC channel in both CNGS experiments,
assuming a detection efficiency of 0.9 and taking into account a
fraction of 0.05 of all neutral current events as background. 
As a matter of fact, the measurement of the atmospheric
parameters also contributes to the $\sin^22\theta_{13}$
sensitivity limit, since correlation effects decrease with a higher
precisions on the atmospheric parameters. Therefore, the
$\sin^22\theta_{13}$ sensitivity at \icarus\ and \opera\ is considerably
improved by including (besides the $\nu_\mu\to\nu_e$
channel) the $\nu_\mu\rightarrow\nu_\tau$ appearance and the
$\nu_\mu$ disappearance channels in the fit.

We have checked for all setups that the results do not depend 
significantly on the energy range, energy resolution, and bin size as 
long as the energy information is sufficient to distinguish the shape of 
the signal from the shape of the background.

The $\sin^22\theta_{13}$ sensitivity of the different experiments is
provided mainly by the information from the $\nu_\mu \rightarrow
\nu_e$ appearance channel. Because of the small value of
$\sin^22\theta_{13}$, the number of $\nu_\mu \rightarrow \nu_e$ CC
events will be very small compared to the $\nu_\mu \rightarrow
\nu_\mu$ CC and NC events.  Furthermore, the events from the intrinsic
$\nu_e$ component of the beam create a background to the oscillation
signal. Thus, in our simulation, we consider as possible backgrounds:
Beam $\nu_e$ CC events, misidentified $\nu_\mu$ CC events,
misidentified $\nu_\tau$ CC events (mainly for CNGS), and
misidentified NC events.  We have calibrated the various background
sources in our simulation carefully with respect to the information
given in the literature. The corresponding references are for \minos\
\Tab~3 of \Ref~\cite{NUMI714}, for \icarus\ \Ref~\cite{Aprili:2002wx},
and for \opera\ \Tab~4 of \Ref~\cite{Komatsu:2002sz}. Using this
information, we can reproduce with high accuracy the numbers of signal
and background events provided by the experimental collaborations,
which can be found in \Tab~\ref{tab:calibration}.

\begin{table}[t!]
\centering
\begin{tabular}{|ll|r|rrrrr|}\hline 
 & & Signal & \multicolumn{5}{c|}{Background}  \\ 
 Experiment & Reference & $\nu_\mu \rightarrow \nu_e$ & $\nu_e \rightarrow \nu_e$ & 
 $\nu_\mu \rightarrow \nu_\mu$ & $\nu_\mu \rightarrow  \nu_\tau$ &
 NC & Total \\  
\hline
\minos & NuMI-L-714~\cite{NUMI714}&8.5&5.6&3.9&3.0&27.2&39.7\\
 \icarus &  T600 proposal~\cite{Aprili:2002wx} & 
51.0 & 79.0 & - & 76.0 & - & 155.0 \\ 
\opera & Komatsu et al.~\cite{Komatsu:2002sz} & 5.8 & 18.0 & 1.0 & 4.6 & 5.2 & 28.8 \\
\hline
\end{tabular}
\mycaption{\label{tab:calibration} The signal and background events
   for the three conventional beam experiments. The reference points
   are $\Delta m^2_{31}=3.0 \cdot 10^{-3} \,\mathrm{eV}^2$,
   $\sin^2\theta_{13}=0.01$ for \minos, $\Delta m^2_{31}=3.5 \cdot
   10^{-3} \,\mathrm{eV}^2$, $\sin^22\theta_{13}=0.058$ for \icarus,
   $\Delta m^2_{31}=2.5 \cdot 10^{-3} \,\mathrm{eV}^2$,
   $\sin^22\theta_{13}=0.058$ for \opera, and $\sin^22\theta_{23}=1$,
   $\Delta m^2_{21}=\sin^22\theta_{12}=\deltacp=0$ in all three
   cases. The nominal exposures are $10\,\mathrm{kt}\,\mathrm{y}$
   (\minos), $20\,\mathrm{kt}\,\mathrm{y}$ (\icarus), and
   $8.25\,\mathrm{kt}\,\mathrm{y}$ (\opera). Note that these numbers are
   different from the ones in \Tab~\ref{tab:chooz}, since different
   reference points and luminosities are used.}
\end{table}

\subsection{Reproduction of the analyses performed by the experimental 
            collaborations}
\label{app:reproduction}

In order to demonstrate the reliability and accuracy of our
calculations, we use in this appendix the analysis techniques from
\Refs~\cite{NUMI714,Aprili:2002wx,Komatsu:2002sz}, and compare our
results with the ones in these references. For this purpose, we 
neglect all correlations and degeneracies, \ie , we set the solar mass
splitting to zero, which also eliminates the solar mixing angle and CP
effects. In addition, we fix the atmospheric mixing angle to $\pi/4$.
Thus, the only remaining parameters are $\theta_{13}$ and $\dm{31}$, 
where $\dm{31}$ is assumed to be exactly known. For both
\minos\ and the CNGS experiments, we use the background uncertainties
given in \Refs~\cite{NUMI714,Aprili:2002wx,Komatsu:2002sz}, \ie , 10\%
for \minos\ and 5\% for \icarus\ and \opera . Then we simulate data
for each value of $\dm{31}$ with $\theta_{13}=0$, and fit these data
with $\theta_{13}$ as the only free parameter. This simplified
procedure leads to a limit similar to $(\sin^22\theta_{13})_\mathrm{eff}$,
which represents the ability to identify a signal (but not to extract
$\stheta$), and is similar to a simple estimate of $S/\sqrt{S+B}$.  

\begin{figure}[t!]
\centering 
   \includegraphics[angle=-90,width=\textwidth]{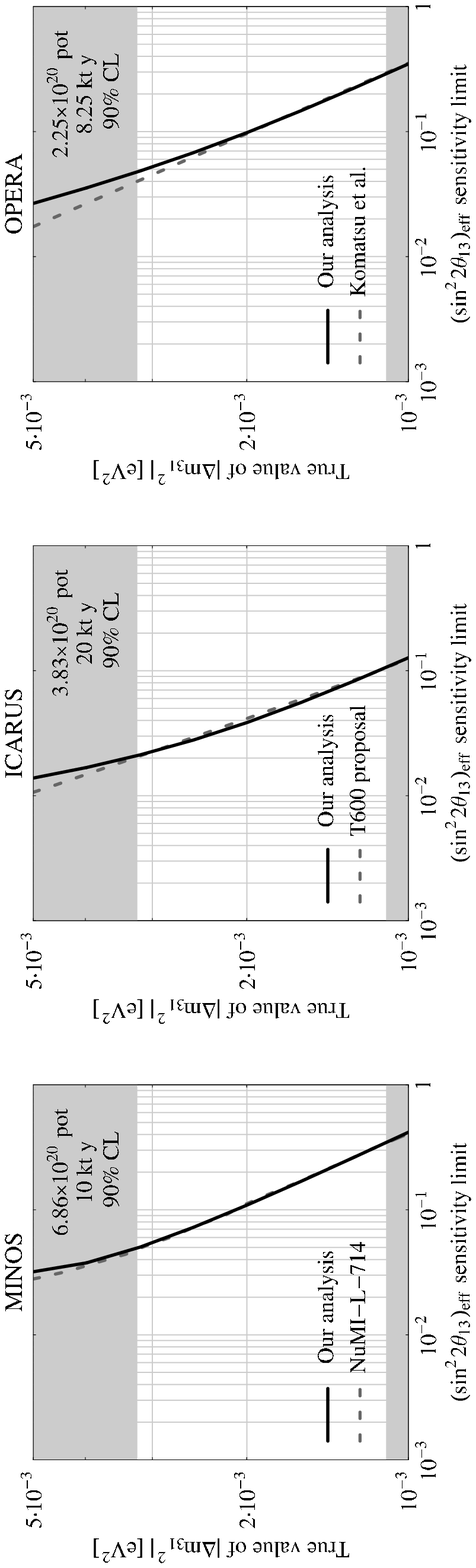} 
   \mycaption{\label{fig:effective} The effective
   $\stheta$ sensitivity limit (as discussed in the main text) at 90\% CL as
   a function of the true value of $\ldm$ for the conventional beam
   experiments. The exposure is $10\,\mathrm{kt}\,\mathrm{y}$ for
   \minos , $20\,\mathrm{kt}\,\mathrm{y}$ for \icarus\ and a nominal
   running time of five years for \opera , corresponding to an
   exposure of $8.25\,\mathrm{kt}\,\mathrm{y}$. The dashed gray curves
   come from the collaborations of the individual experiments and are
   taken from \Ref~\protect\cite{NUMI714} for \minos ,
   \Ref~\protect\cite{Aprili:2002wx} for \icarus , and
   \Ref~\protect\cite{Komatsu:2002sz} for \opera .  The black curves are
   obtained for $\sin^22\theta_{23}=1$ and $\Delta
   m^2_{21}=\sin^22\theta_{12}=\deltacp=0$ for systematics only, \ie,
   correlations and degeneracies are not included, as for the dashed
   curves.} 
\end{figure}

In \figu{effective}, we compare the discussed effective $\stheta$ limit to the ones of the experimental collaborations. The solid black curves represent our
results, whereas the dashed gray curves are taken from
\Refs~\cite{NUMI714,Aprili:2002wx,Komatsu:2002sz}. Within the
Super-Kamiokande allowed atmospheric region, our simulation is in very
good agreement with the results of the different collaborations.  The
slight deviation in the \opera\ curve at large values of $\dm{31}$ 
comes from the efficiencies in \Ref~\cite{Komatsu:2002sz}, since they
are only given as energy-integrated quantities. Thus, it is not
possible to fully reproduce the energy dependence of the events.  This
effect becomes stronger if the oscillation maximum is shifted to
higher energies compared to the reference point used in
\Ref~\cite{Komatsu:2002sz}. However, the influence on our results is
marginal, since \opera\ does not contribute significantly to the
$\theta_{13}$ sensitivity. 

Note that the numbers given in \Tab~\ref{tab:calibration} and the
results shown in \figu{effective} do not allow a comparison of the
different experiment performances, because the reference points used
for the calculation of \Tab~\ref{tab:calibration} are rather
different, and so are the integrated luminosities. For a comparison on
equal footing we refer to \Tab~\ref{tab:chooz} and the discussion in
\Sec~\ref{sec:stheta_conv}.

%%%%%%%%%%%%%%%%%%%%%%%%%%%%%%%%%%%%%
%%           Reactor appendix      %%
%%%%%%%%%%%%%%%%%%%%%%%%%%%%%%%%%%%%%

\section{Simulation of the reactor experiments}
\label{app:reactor}

For our simulation of the reactor neutrino experiments, we closely
follow our previous work in \Ref~\cite{Huber:2003pm}. For the analyses
presented here, we assume a near detector baseline of $L_\mathrm{ND} =
0.15 \, \mathrm{km}$ and we consider two options for the far detector
baseline: $L_\mathrm{FD} = 1.05 \, \mathrm{km}$ corresponds to the
baseline at the CHOOZ site, whereas a baseline $L_\mathrm{FD} = 1.7 \,
\mathrm{km}$ is close to the options considered for several other
sites. We always fix the number of reactor neutrino events in the
near detector to $2.94\cdot 10^6$, which implies that it has the same
size as the far detector at $L_\mathrm{FD} = 1.05 \, \mathrm{km}$ with
$60\,000$ events (assuming the same efficiencies in
both detectors\footnote{Due to a higher background rate in the near
detector, there will more dead time than in the far detector. This
reduces the number of events in the near detector roughly by a factor
of two. We have checked that this has a very small impact on the final
sensitivity.}). We allow an uncertainty of the overall reactor neutrino
flux normalization of $\sigma_\mathrm{abs} = 2.5\%$. For the
normalization error between the two detectors, we use a typical value
of $\sigma_\mathrm{rel} = 0.6\%$. As shown in
\Ref~\cite{Huber:2003pm}, this roughly corresponds to an effective
normalization error of $\sigma_\mathrm{norm} \simeq 0.8\%$.
The total range for the visible energy $E_\mathrm{vis} = E_{\bar\nu} -
\Delta + m_e$ (where $\Delta$ is the neutron-proton mass difference,
and $m_e$ is the electron mass) from 0.5~MeV to 9.2~MeV is divided
into 31 bins.  Furthermore, we assume a Gaussian energy resolution
with $\sigma_\mathrm{res} = 5\% / \sqrt{E_\mathrm{vis}
[\mathrm{MeV}]}$. We remark that our results do not change if a
smaller bin width is chosen, as it would be allowed by the good energy
resolution and the large number of events.
Furthermore, we take into account an uncertainty on the energy scale
calibration $\sigma_\mathrm{cal} = 0.5\%$, and an uncertainty on the
expected energy spectrum shape $\sigma_\mathrm{shape} = 2\%$, which we
assume to be uncorrelated between the energy bins, but fully correlated
between the corresponding bins in near and far detectors (see
\Ref~\cite{Huber:2003pm} for details).  

\begin{figure}[t!]
\centering \includegraphics[width=0.5\textwidth]{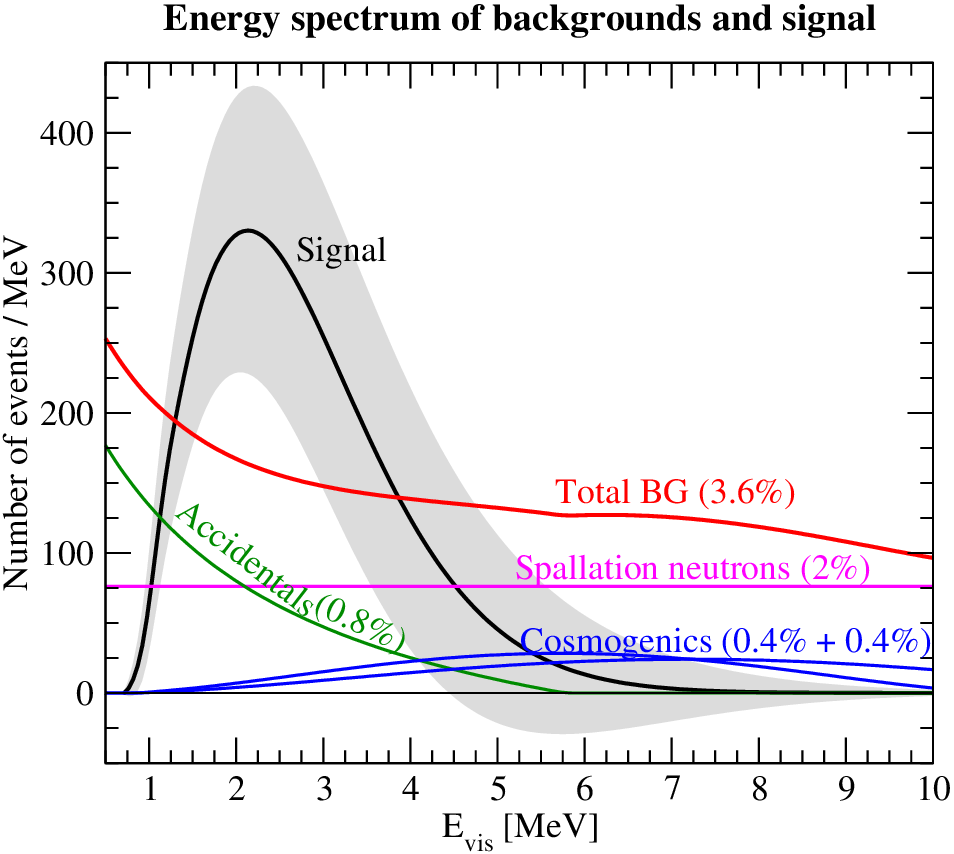}
   \mycaption{\label{fig:BG-spectrum} Energy spectrum of the
   backgrounds from spallation neutrons, accidentals, and cosmogenic
   $^9$Li and $^8$He. The percentage given for each curve corresponds
   to the total number of background events relative to the total
   number of reactor neutrino events for no oscillations.  Also shown
   is the total background spectrum. The curve labeled ``signal''
   corresponds to $N(E_\mathrm{vis}; \stheta=0) - N(E_\mathrm{vis};
   \stheta=0.05)$, where $N(E_\mathrm{vis}; \stheta)$ is the energy
   spectrum for given $\stheta$. The shaded region is the statistical
   error band at 1$\sigma$, \ie, $\pm\sqrt{N(E_\mathrm{vis};
   \stheta=0)}$. Note that the absolute normalizations of the
   backgrounds are exaggerated in this figure.}
\end{figure}

In addition to the analysis as performed in \Ref~\cite{Huber:2003pm},
we have investigated in greater detail the impact of a background for a
reactor experiment of the \DChooz\ type. We take into account four different
background sources with known shape: 
\begin{itemize}
\item
A background from spallation neutrons coming from muons in the rock
close to the detector. This background can be assumed to be flat as a
function of energy to a first approximation (see, \eg, \Fig~48 of
\Ref~\cite{Apollonio:2002gd}).
\item
A background from accidental events. A $\gamma$ from
radioactivity is followed by a second random $\gamma$ with more than
6~MeV faking a neutron signal. Those events are important for low energies.
\item
 Two correlated backgrounds from cosmogenic $^9$Li and $^8$He nuclei.
Both are created by through-going muons and give $\beta$-spectra with
end points of 13.6 and 10.6~MeV, respectively.
\end{itemize}
In \Tab~\ref{tab:backgrounds}, we give for each background a realistic
estimate~\cite{herve} for the expected number of events in near and
far detectors relative to the number of reactor neutrino events.
Since the near detector will have less rock overburden than the far
detector, more background events are expected. They are, however,
compensated by the much larger number of reactor neutrino events in
the near detector. Therefore, we assume to a first approximation the
same {\it relative} sizes for the backgrounds in near and far
detector. In \Fig~\ref{fig:BG-spectrum}, the spectral shape of these
backgrounds is shown. We assume that these shapes are exactly
known, however, the overall normalization of each of the four
background components in the two detectors is allowed to fluctuate
independently with an error of $\sigma_\mathrm{BG} = 50\%$.
In addition to these backgrounds with known shape, we include a
background from an unidentified source with a bin-to-bin
uncorrelated error of 50\%. We assume a
total number of background events of 0.5\%  of the reactor signal in both
detectors, and a flat energy shape for this background.

\begin{table}[b!]
\begin{center}
\begin{tabular}{|llrr|}
\hline
Background type & Spectral shape & BG/Reactor events & $\sigma_\mathrm{BG}$\\
\hline
\multicolumn{4}{|l|}{\bf{Backgrounds with known shape}} \\
\hline
Spallation neutrons  & Flat         & 0.4\%  & 50\% \\
Accidentals          & Low energies & 0.2\%  & 50\% \\
Cosmogenic $^9$Li    & $\beta$-spectrum (end point 13.6~MeV) & 0.2\% & 50\%\\
Cosmogenic $^8$He    & $\beta$-spectrum (end point 10.6~MeV) & 0.2\% & 50\%\\
\hline
\multicolumn{2}{|l}{Bin-to-bin correlated BG total:} & 1.0\%  & \\
\hline
\hline
\multicolumn{4}{|l|}{\bf{Bin-to-bin uncorrelated background}} \\
\hline
Unknown source & Flat &  0.5\% & 50\% \\   
\hline
\end{tabular}
\end{center}
\mycaption{\label{tab:backgrounds} Backgrounds included in our reactor
  experiment analysis. For each background source, the column
  ``BG/Reactor events'' refers to the total number of background
  events in the energy range between 0.5 and 9.2~$\mathrm{MeV}$
  relative to the total number of reactor neutrino events for no
  oscillations~\protect\cite{herve}. We assume the same magnitudes of the
  backgrounds relative to the total events in the near and far
  detectors. For the backgrounds with known shape,
  $\sigma_\mathrm{BG}$ is the uncertainty of the overall
  normalization. For the uncorrelated background, $\sigma_\mathrm{BG}$
  is the error on the number of events in each bin, which is
  uncorrelated between different bins (31~bins). All backgrounds are
  uncorrelated between the two detectors.}
\end{table}

As a general trend, we find that backgrounds with known shape do not
significantly affect the sensitivity. This holds independently of the
integrated luminosity. Even increasing the numbers given in
\Tab~\ref{tab:backgrounds} by a factor five does not change the
picture. This behavior can be understood in terms of
\Fig~\ref{fig:BG-spectrum}, where we show the signal (the spectrum
without oscillation minus the spectrum for $\stheta = 0.05$) and its
statistical error compared to the various background spectra. For
illustration, background levels significantly larger than in
\Tab~\ref{tab:backgrounds} are assumed. From this figure, it is
obvious that the spectral shape of the signal is very different from
that of all of the background components. Already at modest
luminosities, such as for \DChooz , enough spectral information is
available to determine the backgrounds with sufficient accuracy, which
means that it is not possible to fake the signal within the
statistical error by fluctuations of the background components. In
contrast, we find that the bin-to-bin uncorrelated background only
plays a minor role for experiments of the size of \DChooz , whereas it
becomes important for large experiments, such as \ReactorII. In the
latter case, a background level of 0.5\% with a bin-to-bin
uncorrelated error larger than about 20\% would significantly affect
the sensitivity. We conclude that for large reactor experiments, the
shape of the expected background has to be well under
control.\footnote{Note that the above statements quantitatively depend
to some extent on the chosen number of bins. For example, a bin-to-bin
uncorrelated error of 20\% for 10 bins has, in general, a different
impact than such an error for 60 bins.  This can be understood by the
fact that for a large number of bins, the simultaneous fluctuation of
neighboring bins becomes unlikely. The same argument holds for the
flux shape uncertainty $\sigma_\mathrm{shape}$.}

In \Tab~\ref{tab:reactor-setups}, we summarize the experimental
parameters which we use for the two setups \DChooz\ and \ReactorII\ in
the main text of this study. For the \DChooz\ setup, we stick closely
to the configuration discussed in \Ref~\cite{doubleChooz}. We have
checked that the effect of the slightly different distances ($1.0 \,
\mathrm{km}$ and $1.1 \, \mathrm{km}$) of the far detector location
from the two different reactor cores at the CHOOZ site is very
small. Hence it is a good approximation to consider the average
baseline of $1.05 \, \mathrm{km}$ for both cores. In order to obtain
robust results, we include all systematical errors as well as
backgrounds.

\begin{table}[t!]
\begin{center}
\begin{tabular}{|l|r@{\qquad}r|}
\hline
& \DChooz & \ReactorII \\
\hline
Luminosity ${\cal L}$ 
& $288 \,\mathrm{t\, GW \,y}$ & $8000 \,\mathrm{t\, GW \,y}$ \\
Number of events in ND & $2.94 \cdot 10^6$ & $2.94 \cdot 10^6$ \\
Number of events in FD & $6 \cdot 10^4$ & $6.36 \cdot 10^5$ \\
Near detector baseline & $0.15 \,\mathrm{km}$  & $0.15 \,\mathrm{km}$ \\
Far detector baseline & $1.05 \,\mathrm{km}$  & $1.70 \,\mathrm{km}$\\
Energy resolution & \multicolumn{2}{r|}{
$ \sigma_\mathrm{res} = 5\%/\sqrt{E [\mathrm{MeV}]}$}\\
Visible energy range &
\multicolumn{2}{r|}{$0.5 - 9.2 \,\mathrm{MeV}$ (31 bins)}\\  
Individual detector normalization & 
$\sigma_\mathrm{rel} = 0.6\%$ & $\sigma_\mathrm{rel} = 0.6\%$ \\
Flux normalization & 
$\sigma_\mathrm{abs} = 2.5\%$ & $\sigma_\mathrm{abs} = 2.5\%$ \\
Flux shape uncertainty & 
$\sigma_\mathrm{shape} = 2\%$ & $\sigma_\mathrm{shape} = 0$ \\
Energy scale error & 
$\sigma_\mathrm{cal} = 0.5\%$ & $\sigma_\mathrm{cal} = 0$ \\
Backgrounds included & Yes & No\\
\hline
\end{tabular}
\end{center}
\mycaption{\label{tab:reactor-setups} Characteristics of the two
  reactor experiments \DChooz\ and \ReactorII.}
\end{table}

The large reactor experiment \ReactorII\ corresponds to the same setup
as already used in \Ref~\cite{Huber:2003pm}. It represents an ideal
configuration without backgrounds and any systematical errors beyond
overall normalization errors. This setup has been chosen to
illustrate how an optimal reactor experiment would fit into the
general picture of the next ten years of oscillation physics and,
to obtain CP-complementary information. Several
proposals which are close to our \ReactorII\ setup are currently
discussed~\cite{whitepaper}. 

\section{The definition of the $\boldsymbol{\stheta}$ sensitivity limit}
\label{app:stheta}

In this appendix, we discuss the definition of the
$\stheta$ sensitivity limit. Although this definition is very general,
we mainly focus on the $\nu_\mu\to\nu_e$ appearance
channel, since one has to deal extensively with parameter
correlations and degenerate solutions in this case.  Let us first define our
$\stheta$ sensitivity and then discuss its properties.

\begin{definition}
\label{def:sl}
   We define the {\bf $\boldsymbol{\stheta}$ sensitivity limit} as the
   largest value of $\stheta$, which fits the true value $\stheta=0$
   at the chosen confidence level. The largest value of $\stheta$ is
   obtained from the projections of all (disconnected) fit manifolds (best-fit manifold and degeneracies) onto the $\stheta$-axis.
\end{definition}

\begin{figure}[t!]
\begin{center}
\includegraphics[width=8cm]{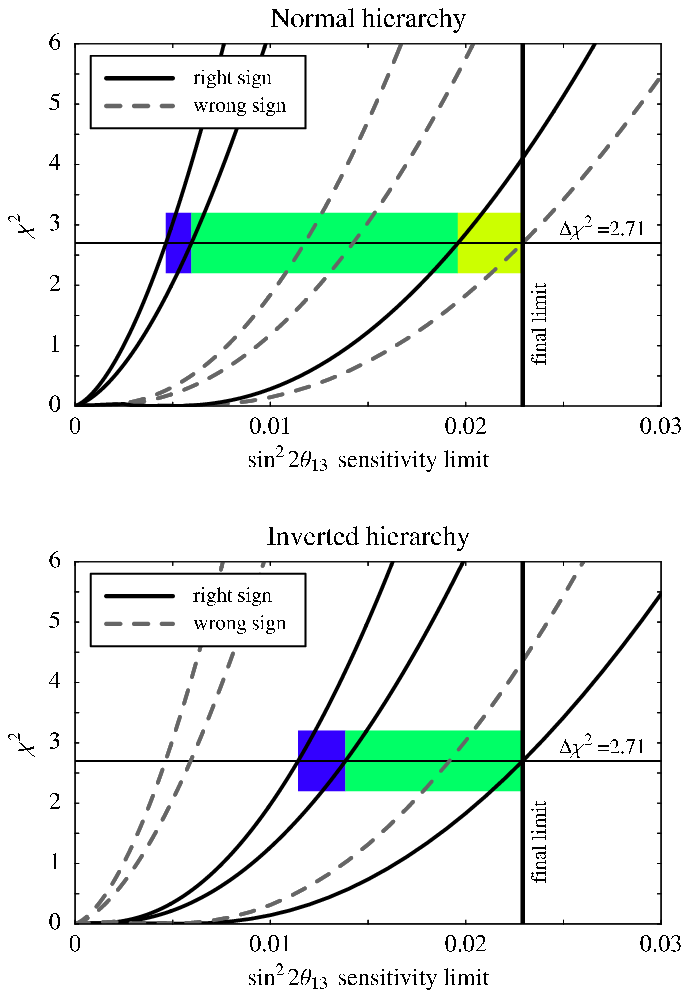}
\end{center}
\mycaption{\label{fig:did1} The $\chi^2$ as function of $\stheta$ for
   \JHFSK. For the true values of the oscillation parameters, we
   choose the current best-fit values from \eq~(\ref{eq:bfp}),
   $\stheta=0$, $\deltacp=0$ (for curves without correlations only),
   and normal (upper plot) or inverted (lower plot) mass
   hierarchies. The solid curves in each plot are obtained by fitting
   with the same mass hierarchy as has been used to calculate the
   reference rate vector (``right-sign''), whereas for the dashed
   curves the wrong mass hierarchy has been used
   (``wrong-sign''). Within each group of solid or dashed curves, the
   left curve determines the statistics-only limit, the middle curve
   the statistics+systematics limit, and the right curve the
   statistics+systematics+correlations limit. Note that the wrong-sign
   minimum has not exactly the same position in parameter space as the
   original minimum.}
\end{figure}

Since for future experiments no data are available, one has to simulate
data by calculating a ``reference rate vector'' for a fixed set of
``true'' parameter values. In general, the experiment performance
depends on the chosen set of true parameter values, and it is interesting to
discuss this dependency in many cases. It is especially relevant for
the true values of $\sdm$ and $\ldm$, which we usually choose within
their currently allowed ranges. According to Definition~\ref{def:sl}, 
we choose the true value $\stheta = 0$ to
calculate the $\stheta$ limit, since we are interested in the bound 
on $\stheta$ if no positive signal is observed. 
Moreover, this choice has the following
advantages:
\begin{itemize}
\item
Since for $\stheta = 0$ the phase $\deltacp$ becomes unphysical, the
sensitivity limit will be {\it independent of the true value of $\deltacp$}.
\item
For $\stheta = 0$, the reference rate vectors for the normal and the
inverted mass hierarchies are approximately equal, which implies that the
sensitivity limit {\it hardly depends on the true sign
of $\ldm$} (see also the discussion related to \Fig~\ref{fig:did1} later).
\end{itemize}

Once the reference rate vector has been obtained, the fit manifold in
the six-dimensional space of the oscillation parameters is given by
the requirement $\Delta \chi^2 \le \mathrm{CL}$ (\eg, at the
90\% confidence level, we have $\mathrm{CL} = 2.71$ for 1 d.o.f.). In addition
to the allowed region which contains the best-fit point (``best-fit
manifold''), one or more disconnected regions (``degenerate
solutions'') will exist, and each of them may have a rather
complicated shape in the six-dimensional space (``correlations''). The
final sensitivity is given by the largest value of $\stheta$ which fits
$\stheta=0$. It is obtained by projecting {\it all} these disconnected fit regions onto the $\stheta$-axis, where the projection takes into account the correlations. Hence, this procedure provides a straightforward
method to take into account correlations and degeneracies.
Thus, for the case of the $\nu_\mu\to\nu_e$ appearance channel, our
definition of the $\stheta$ sensitivity limit includes the intrinsic
structure of \equ{beam}. This equation reflects that an appearance experiment
is only sensitive to a particular combination of parameters.  The projection onto the $\stheta$-axis takes into account that all the other parameters can be only
measured with a certain accuracy by the experiment itself. Moreover, in complicated
cases (\eg, for neutrino factories) local minima may appear in the
projection of the $\chi^2$-function onto the $\stheta$-axis, and the
$\chi^2$-function can intersect the chosen confidence level multiple
times. In this case, we choose by definition the rightmost of these
intersections. Hence, the sensitivity limit, as defined above, refers to the
potential of an experiment (or combination of experiments) to extract
the value of the parameter $\stheta$ from \equ{beam} convolved
with all the simulation information. 

In this study, we compare the $\stheta$ sensitivity limit of a given
experiment to a so-called $(\sin^2 2 \theta_{13})_{\mathrm{eff}}$
sensitivity limit in some cases. The $(\sin^2 2 \theta_{13})_{\mathrm{eff}}$
sensitivity limit roughly corresponds to the
potential of a given experiment to observe a positive signal, which is
parameterized by some (unphysical) mixing parameter $(\sin^2 2
\theta_{13})_{\mathrm{eff}}$:
\begin{definition}
   The {\bf $\boldsymbol{(\stheta)_{\mathrm{eff}}}$ sensitivity limit}
   is defined as the sensitivity limit from statistics and systematics
   only which is computed for $\deltacp=0$ by fixing all other
   oscillation parameters to their true values.
\end{definition}

In order to illustrate the impact of systematics, correlations, and
degeneracies, we often use ``bar charts'' (see, for example,
\Figs~\ref{fig:interncomp} and \ref{fig:externcomp}), where the final
$\stheta$ sensitivity is obtained by successively switching on
systematics, correlations, and degeneracies.  In these bar charts, the
statistics-only $\stheta$ sensitivity (left edge of the bar) is
computed for all oscillation parameters fixed and $\deltacp=0$, the
statistics+systematics sensitivity limit corresponds to the
$(\stheta)_{\mathrm{eff}}$ sensitivity limit, the
statistics+systematics+cor\-re\-lations limit corresponds to the
sensitivity limit for the best-fit manifold only (no degenerate
solutions included), and the final sensitivity limit (right edge of
the bar) corresponds to Definition~\ref{def:sl}.
 
In the following, we illustrate in greater detail how the
$\stheta$ limit is obtained, how the bar charts are constructed, and 
how the $\stheta$ and $(\stheta)_{\mathrm{eff}}$
limits are related to each other at the example of the \JHFSK\ experiment. 
We focus mainly on the $\mathrm{sgn}(\ldm)$-degeneracy and the correlation
between $\theta_{13}$ and $\deltacp$, which is of particular relevance for the $\nu_\mu\to\nu_e$ appearance channel at superbeams.

In \figu{did1}, the $\chi^2$ is shown as a function of $\stheta$ for
the ``right-sign'' and ``wrong-sign'' solutions, where in the upper
(lower) plot the normal (inverted) hierarchy has been chosen to
calculate the reference rate vector.  The right-sign solution is
obtained by fitting with the same sign of $\ldm$ as the reference rate
vector has been calculated with, \ie, the ``right'' neutrino mass
hierarchy is used, whereas the wrong-sign solution is obtained by
fitting with the opposite sign of $\ldm$, \ie, the ``wrong'' mass
hierarchy is used.  The different curves in each group with the same
curve style correspond, from the left to the right, to the
statistics-only, statistics+systematics, and
statistics+systematics+correlations sensitivity limits, where these
limits are obtained from the intersection of the $\chi^2$ with the
$\Delta\chi^2=2.71$ line. The bar charts are constructed from the
corresponding curves, as one can easily read off the figure.

Comparing the normal and inverted mass hierarchy plots in \figu{did1},
one can observe a symmetry between the right- and wrong-sign
solutions: The curves for the normal mass hierarchy and $\ldm>0$
(right sign) are very similar to the ones of the inverted mass
hierarchy and $\ldm>0$ (wrong sign). This can be understood in terms
of the identical appearance rate vectors for the normal and inverted
mass hierarchies for the true value of $\stheta=0$. However, since the
role of the $\ldm>0$ curves is different for the normal and inverted
mass hierarchies, \ie, they either correspond to the best-fit manifold
(right sign) or the $\mathrm{sgn}(\ldm)$-degeneracy (wrong sign), the
bar charts are, by definition, very different, since they are
originally determined by the best-fit solution. However, one can
easily see that the final sensitivity limit does not depend on the
mass hierarchy~\cite{Huber:2002rs}. This property comes from the fact
that the degeneracy part does not contribute to the final sensitivity
if the best-fit $\stheta$ sensitivity is already worse than the
degenerate solution sensitivity. Since there is hardly a difference
between final sensitivity limits for the different mass hierarchies,
we usually show the normal mass hierarchy sensitivity limit. In fact,
there is a small difference between the final sensitivity limits for
the different mass hierarchies, which mainly comes from the
disappearance channels.

\begin{figure}[t!]
\begin{center}
\includegraphics[width=8cm]{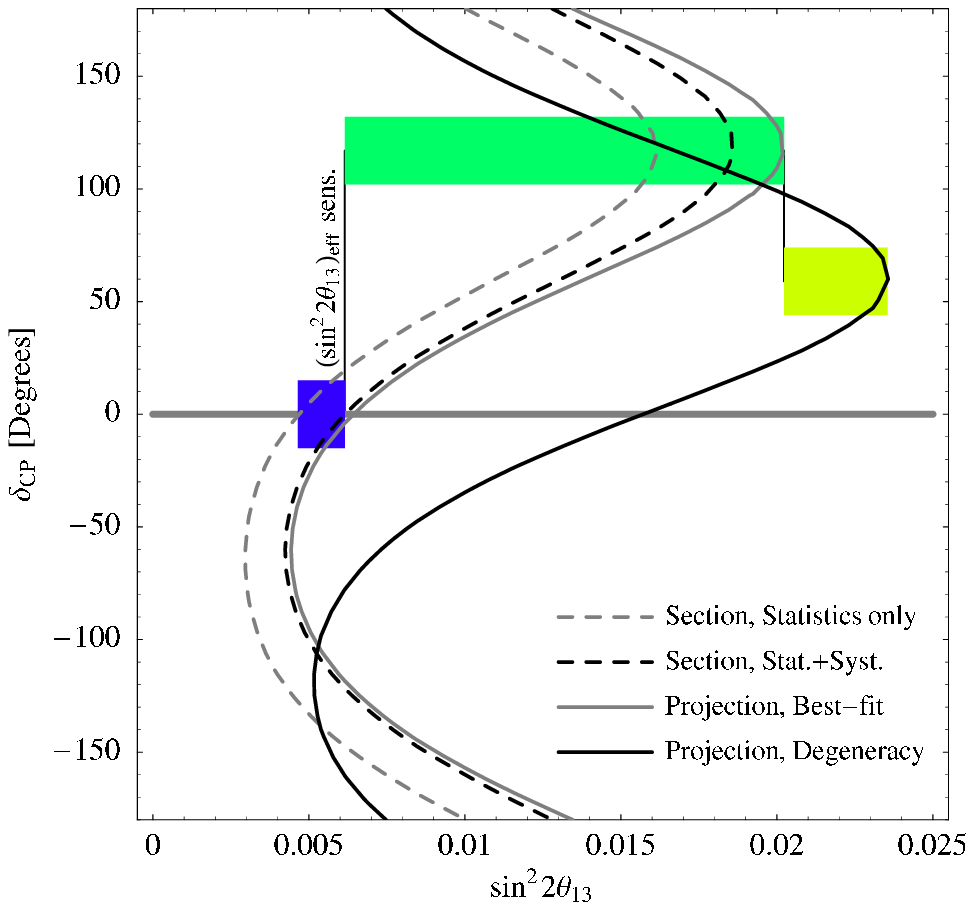}
\end{center}
\mycaption{\label{fig:did2} The 90\% CL fit manifold (1 d.o.f.) in the
   $\stheta$-$\deltacp$-plane for \JHFSK.  For the true values of the
   oscillation parameters, we choose the current best-fit values from
   \eq~(\ref{eq:bfp}) and $\stheta=0$. The different curves correspond
   to various sections (un-displayed oscillation parameters fixed) and
   projections (minimized over un-displayed oscillation parameters) as
   described in the plot legend. The bars demonstrate the individual
   contributions to the final $\stheta$ sensitivity limit. Note that
   for the sgn($\ldm$)-degenerate solution, we only show the final
   projection.  }
\end{figure}

Let us now illustrate the impact of the correlation between $\stheta$
and $\deltacp$. Therefore, we show in \figu{did2} the fit manifold
in the $\stheta$-$\deltacp$-plane.  The $\stheta$ sensitivity limit
is again obtained from the projection onto the $\stheta$-axis.  In
\figu{did2}, the individual contributions to the bar chart are
illustrated by showing different sections (un-displayed oscillation
parameters fixed, \ie, no correlations) and projections (minimized
over un-displayed oscillation parameters, \ie, they include correlations) of
the fit manifold. One can see that both edges of the leftmost (blue)
bar are computed for $\deltacp=0$. This illustrates that if $\deltacp$
and all the other oscillation parameters except $\stheta$ are fixed at
the true values, much stronger bounds on $\stheta$ can be obtained,
corresponding to the $(\stheta)_\mathrm{eff}$ limit. The limit gets
considerably weaker if the $\chi^2$-function is minimized over
$\deltacp$, as well as all oscillation parameters which are not shown,
which leads to the ``correlation bar''. In fact, from \figu{did2}, one
can see that the largest part of the correlation bar comes from the
correlation with $\deltacp$~\cite{Kajita:2001sb}, and only the small difference between the dark dashed and the light solid curves comes from the correlation with
the other oscillation parameters. The final sensitivity limit is then
obtained as the maximum value of $\stheta$ which fits $\stheta=0$ 
including all degenerate solutions.

\end{appendix}

%%%%%%%%%%%%%%%%%%%%%%%%%%%%%%%%%%%%%%%%%%%%%%%%%%%%%%%%%%%%%%%%%%%%%%
%%%%%%%%%%             References                         %%%%%%%%%%%%
%%%%%%%%%%%%%%%%%%%%%%%%%%%%%%%%%%%%%%%%%%%%%%%%%%%%%%%%%%%%%%%%%%%%%% 

\end{document}